\documentclass[aps,prd,nofootinbib,twocolumn,superscriptaddress,floatfix,notitlepage]{revtex4-1}
\pdfoutput=1

\usepackage{ulem}
\usepackage{graphicx}
\usepackage{amsmath,amsfonts,amssymb}
\usepackage[dvipsnames]{xcolor}
\usepackage[breaklinks,colorlinks,urlcolor=blue,citecolor=blue,linkcolor=magenta]{hyperref}
\usepackage{enumitem}
\usepackage{slashed}
\usepackage{bm}
\usepackage{stackrel}

\usepackage{tikz}
\usepackage{tkz-euclide}
\usetikzlibrary{shapes.misc}
\usetikzlibrary{decorations.pathmorphing}	
\tikzset{
    v/.style={decorate, decoration={snake, segment length=3mm, amplitude=0.75mm}, draw},
    f/.style={draw=black, postaction={decorate},
        decoration={markings,mark=at position .6 with {\arrow[very thick]{latex}}}},
    fb/.style={draw=black, postaction={decorate},
        decoration={markings,mark=at position .4 with {\arrowreversed[very thick]{latex}}}},
    fnar/.style={draw=black},
    g/.style={decorate, draw=black,
        decoration={coil,amplitude=3pt, segment length=3.5pt}},
    s/.style={dashed,draw=black, postaction={decorate},
        decoration={markings,mark=at position .55 with {\arrow[very thick]{latex}}}},
    sb/.style={dashed,draw=black, postaction={decorate},
        decoration={markings,mark=at position .55 with {\arrowreversed[draw=black,very thick]{latex}}}},
    snar/.style={dashed,draw=black,line width =1.25pt},
    cross/.style={cross out, draw=black, minimum size=2*(#1-\pgflinewidth), inner sep=0pt, outer sep=0pt},
cross/.default={3pt},
}

\newcommand{\be}{\begin{equation}} 
\newcommand{\ee}{\end{equation}}
\newcommand{\al}[1]{\begin{align}\begin{aligned} #1 \end{aligned}\end{align}}
\newcommand{\ket}[1]{|#1\rangle}
\newcommand{\bra}[1]{\langle #1 |}
\newcommand{\braket}[1]{\langle #1 \rangle}
\newcommand{\dd}{\text{d}}
\newcommand{\pdag}{\phantom{\dagger}}
\newcommand{\ptwo}{\phantom{2}}
\newcommand{\pstar}{\phantom{\star}}

\usepackage{comment}

\begin{document}
\begin{minipage}{8cm}
\vspace{-1cm}
    \begin{flushright}
    DESY-25-183
\end{flushright}
\end{minipage}

\title{The effects of non Bunch--Davies initial conditions on gravitationally produced relics}
\author{Enrico Bertuzzo}
\email{enrico.bertuzzo@unimore.it}
\affiliation{Dipartimento di Scienze Fisiche, Informatiche e Matematiche, Università degli Studi di Modena
e Reggio Emilia, Via Campi 213/A, I-41125 Modena, Italy}
\affiliation{INFN sezione di Bologna, via Irnerio 46, 40126 Bologna, Italy}

\author{Gabriel M. Salla}
\email{gabriel.massoni.salla@desy.de}
\affiliation{Deutsches Elektronen-Synchrotron DESY, Notkestr. 85, 22607 Hamburg, Germany}
\affiliation{Instituto de Física, Universidade de São Paulo, C.P. 66.318, 05315-970 São Paulo, Brazil}

\author{Andrea Tesi}
\email{andrea.tesi@fi.infn.it}
\affiliation{INFN Sezione di Firenze, Via G. Sansone 1, I-50019 Sesto Fiorentino, Italy
}

\date{\today}

\begin{abstract}
    Typical gravitational production of relics from amplification of inflationary perturbations assumes Bunch--Davies initial conditions, i.e. a vacuum with initially no particles. In this paper we investigate the impact of non Bunch--Davies initial conditions to the final abundance of relics, with particular attention to the parameter space where the total dark matter abundance is reproduced. We present a general framework for any initial condition, through which we show their non-trivial effect on both spectrum and late-time abundance. We argue that for particles whose source of conformal symmetry breaking comes only from a mass term (spin-1/2 fermions and conformally coupled scalars), the choice of initial conditions has little impact on the mass range relevant to dark matter. For other particles, e.g. the longitudinal mode of spin-1, we see a deviation from the standard computation. We exemplify and quantify our results with an initial thermal state and a two-stage inflation scenario, highlighting that the total dark matter can be obtained for a wide range of masses.
\end{abstract}

\maketitle

\section{Introduction}
The persistent lack of non-gravitational signals in our quest for Dark Matter (DM) may be a strong indicator that DM interacts with the Standard Model (SM) only via gravity. If this is the case, then the first question we need to address is: how was the DM abundance generated? It is well known that gravitational interactions are, by far, too feeble to allow such DM candidate to come in thermal contact with the SM bath. However, other mechanisms, generically dubbed as ``gravitational production'' are at play in the early universe and may generate the desired abundance. Here we focus on one such mechanisms, cosmological Gravitational Particle Production (GPP), that consists in the production of particles from amplification of quantum fluctuations in the evolving cosmological background. This phenomenon, known since the sixties\,\cite{Parker:1968mv,Parker:1969au,Parker:1971pt,Parker:2025jef}, is completely analogous to the generation of perturbations in the usual inflationary theory. The standard picture can be summarized as follows: suppose the system is in the ground state of the field of interest, $\ket{0_\text{in}}$ (usually called Bunch--Davies vacuum), at the beginning of inflation. In the usual picture, the Bunch--Davies vacuum corresponds to the absence of particles. In Heisenberg picture, the system remains in this state, while the Hamiltonian inherits a time dependence from the evolution of the cosmological background, causing the ground state $\ket{0_\text{out}}$ at later times to not coincide, in general, with $\ket{0_\text{in}}$. An observer at late times will interpret the $\ket{0_\text{in}} \neq \ket{0_\text{out}}$ mismatch as the presence of quanta of the field under discussion: particles have been created during the cosmological evolution. Another way to see this is that the effective action $S_{\rm eff}$ for the evolving sector, $e^{i S_{\rm eff}}=\langle 0_{\rm out}|0_{\rm in}\rangle$, develops an imaginary part.

Recent years have seen a resurgence of interest in GPP as a way to produce DM mainly for two reasons: first, because of the lack of non-gravitational signals mentioned above; second, because GPP is inevitable and must be quantified, for it generates a population that may serve as initial condition for any subsequent production mechanism that does not erase initial conditions (like freeze-in, for instance). More specifically, the spectrum and abundance of particles of spin-0\,\cite{Parker:1969au,Lyth:1996yj,Chung:1998zb,Kolb:1998ki,Chung:2001cb,Chung:2004nh,Chung:2018ayg,Ema:2018ucl,Basso:2021whd,Kolb:2022eyn,Redi:2022myr,Jenks:2024fiu,Racco:2024aac,Belfiglio:2024xqt,Verner:2024agh,Garcia:2025rut}, spin-1/2\,\cite{Parker:1971pt,Lyth:1996yj,Chung:2011ck,Adshead:2015kza,Ema:2019yrd,Redi:2022myr,Koutroulis:2023fgp}, spin-1\,\cite{Graham:2015rva,Ema:2019yrd,Kolb:2020fwh,Ahmed:2020fhc,Arvanitaki:2021qlj,Redi:2022zkt,Capanelli:2024pzd,Capanelli:2024rlk}, spin-3/2\,\cite{Kallosh:1999jj,Giudice:1999yt,Kolb:2021xfn,Kaneta:2023uwi}, spin-2\,\cite{Kolb:2023dzp} and higher-spin\,\cite{Alexander:2020gmv} particles have been studied in detail (see also the reviews\,\cite{Ford:2021syk,Kolb:2023ydq} and references therein). 
GPP generates energy density with an inevitable component of isocurvature perturbations \cite{Chung:2004nh,Garcia:2023qab}, whose importance on cosmologically large scales depends on the details of the models. In all the cases where the bounds are evaded, it can be considered a signature of this production mechanism, together with the gravitational waves generated by such isocurvature perturbations\,\cite{Ebadi:2023xhq,Garcia:2025yit}. The effects of a non-standard cosmological evolution have also been studied\,\cite{Bertuzzo:2024fns}. Three equivalent approaches have been used to compute observables:\footnote{Comparison among these methods and further discussion can be found in refs.\,\cite{Kaneta:2022gug,Chakraborty:2025zgx,Feiteira:2025rpe}.}
(i) the Bogoliubov approach, in which time evolution is described via a Bogoliubov transformation (see references above); (ii) the stochastic approach, in which a Fokker--Planck equation is solved to compute the probability of finding a certain configuration of the field\,\cite{Markkanen:2018gcw,Padilla:2019fju,Cosme:2020nac,Feiteira:2025rpe}; (iii) the effective action approach, in which the overlap $|\braket{0_\text{out}|0_\text{in}}|^2$ is computed using an effective action\,\cite{Garani:2024isu,Garani:2025qnm,Alexeev:2025tpj}. In this work, our computations will be mostly done using Bogoliubov transformations. However, for production that takes place during inflation, we also compute GPP reusing the standard inflationary power spectrum derivation.

The main aim of this work is to start exploring what happens when one of the assumptions of the standard computation is relaxed. Namely, we are not going to take a Bunch--Davies vacuum $\ket{0_\text{in}}$ at the beginning of ``visible inflation'' (i.e. the last 60 $e$-folds or so of inflation), but instead assume the existence of some pre-inflationary dynamics that generates a state $\ket{\psi}$ populated by particles. This problem is not new in the context of inflationary perturbations and was extensively studied in the literature (see refs.\,\cite{Gasperini:1993yf,Kaloper:2002cs,Danielsson:2002kx,Burgess:2002ub,Holman:2007na,Wang:2007ws,Agarwal:2012mq,Aravind:2013lra,Flauger:2013hra,Danielsson:2018qpa,Bhattacharya:2005wn} for a partial list of references)
. Nevertheless, to the best of our knowledge, the production of non-interacting relics, e.g. DM, in the presence of initial conditions differing from the Bunch--Davies one was only mentioned but not quantified\,\cite{Parker:1968mv,Parker:1969au,Parker:1971pt,Agullo:2011xv,Agullo:2010ws}. We thus propose to fill this gap and discuss in depth the effects of non Bunch--Davies initial conditions on the final spectrum and abundance of relics, in particular DM. Naively, one would expect the final number of gravitationally produced particles to simply be the sum between the particles present at the beginning of the evolution and those produced by GPP. As we are going to show in detail, this is not the case.

To exemplify our arguments, we concretely consider two different scenarios. The first consists of the field under consideration thermalizing and reaching thermal equilibrium before visible inflation, remaining agnostic on the precise dynamics that generates this initial condition. The second consists in a two-stage inflationary scenario, that, for concreteness, we will take to be two de Sitter stages with an intermediate radiation domination phase.
As we are going to see, depending on the case considered, the final spectrum and abundance can be drastically different than in the standard case. We will also comment on how to compute the final abundance for more general initial states.

The paper is organized as follows: in sec.\,\ref{sec:review} we discuss the formalism that allows to compute the final particle spectrum and abundance of gravitationally produced particles. We keep our discussion general, and show our results in terms of a generic initial state $\ket{\psi}$. In sec.\,\ref{sec:results} we present our main results focusing on two examples: one in which we assume thermal initial conditions without specifying its dynamical origin (sec.\,\ref{sec:thermal}) and one in which we follow a two-stage inflationary evolution in which the universe evolves through a quasi de Sitter and radiation phase before entering the last stage of inflation (sec.\,\ref{sec:pre_inflation}). In sec.\,\ref{sec:conclusions} we draw our conclusions. We also add a number of appendices: in app.\,\ref{app:conventions_and_fields} we discuss the conventions we adopt for spin-0, 1/2 and 1 fields. In app. \ref{app:coherent} we discuss why coherent initial states do not modify GPP and their difference between thermal and squeezed states. We present in app.\,\ref{app:spectrum_two_stage_inflation} approximate analytical results for the spectrum and number density of the two-stage inflation scenario discussed in sec.\,\ref{sec:pre_inflation}. In app.\,\ref{app:bounds} we instead discuss different limits on the parameter space (isocurvature, Lyman-$\alpha$ and white noise). Throughout the manuscript, we use natural units $\hbar = c = 1$ and write the Friedmann--Lemaître--Robertson--Walker (FLRW) metric as $ds^2 =a(\eta)^2( d\eta^2 - d\bm{x}^2)$, with $\eta$ denoting conformal time and bold quantities three-vectors. A prime represents the derivative with respect to $\eta$, $X'(\eta) \equiv dX(\eta)/d\eta$. Other useful conventions are summarized in app.\,\ref{app:conventions_and_fields}.


\section{GPP for generic initial states}\label{sec:review}
We now review how to compute the particle spectrum and total abundance of gravitationally produced massive particles in an expanding cosmological background. We will keep our discussion generic, so that it can be easily adapted to any initial state. The computation can be done either in terms of mode functions and Bogoliubov transformations or directly in terms of the field power spectrum. In this work we compute everything with fields rescaled by their Weyl factor $\kappa$. Any (massive) field $X$, after a Weyl rescaling, $\tilde X\equiv a(\eta)^\kappa X$, satisfies the quantum equation $$\Box \tilde X+m^2 a^2 \tilde X + (\text{curvature})\tilde X=0,$$
where $\Box\equiv \partial_\eta^2-\partial_{\vec x}^2$ is the flat space D'Alembertian. Scalars have $\kappa=1$, fermions $\kappa=3/2$ and gauge field $\kappa=0$. The curvature term disappears when $X$ is conformally coupled to gravity, which is the case for fermions and massless gauge fields. Instead, light scalars (Goldstone bosons) are naturally minimally coupled to gravity. Notice that since quantum commutation relations are defined at equal times, the Weyl-rescaled $\tilde X$ obeys the same quantization as its parent field $X$, modulo an overall normalization that we fix later.

In this work we want to compute the energy density of the field $X$, generated by GPP, i.e. we are interested in
\begin{equation}
   \lim_{\eta\to+\infty} \langle T^{00}_{X}(\eta)\rangle \equiv \langle \rho_X(\eta)\rangle\approx m \langle n_X(\eta)\rangle\,,
\end{equation}
where in the last equality we have used the fact that, at late times, the DM is collisionless and non-relativistic. As such, eventually, we are most primarily interested in the calculation of $\langle n_X\rangle$. In general, however, the relevant phenomenological quantity is $\rho_X$, which can be computed in terms of the energy density of the Weyl-rescaled field $\tilde X$ as
\begin{equation}\label{eq:energy_density}
    a^4 \langle\rho_X(\eta)\rangle = \langle\mathcal{H}(\eta)\rangle/V\,,
\end{equation}
where $\mathcal{H}/V$ is the energy density of the Weyl-rescaled field $\tilde X$. 

We will discuss two equivalent approaches used  in the calculation of eq.~\eqref{eq:energy_density}, for the interesting cases of scalars and Majorana fermions, $\tilde X=\{\phi,\chi\}$. Notice that this includes also the longitudinal component of a massive vector, which naturally appears as a minimally coupled scalar, and on which we will focus in subsequent sections. Using these two simple cases will allow us to explain the effect that we want to track. More detailed calculations are found in app. \ref{app:conventions_and_fields}.

\subsection{Mode functions and Bogoliubov transformations}\label{sec:mode_functions_Bogoliubov}
In this subsection we discuss the derivation of $\mathcal{H}$ using Bogoliubov transformations, which conventionally relate two sets of creation/annihilation operators acting on the Fock space of $\tilde X$ in the far past and far future. We consider here the case of $\tilde X=\{\phi,\chi\}$, free fields in FRW with mass $m$, which can be expanded as follows 
\begin{eqnarray}\label{eq:quantum_field}
\phi(x) & = & \int \frac{d^3k}{(2\pi)^3} \left[a_{k}^{\pdag}\,u_{k}^{\pdag}(\eta) + a_{-k}^\dag u_{-k}^{\star\pdag}(\eta)\right] e^{i \bm{k}\cdot \bm{x}}\,,\nonumber\\
\chi(x) &=& \int \frac{d^3k}{(2\pi)^3} \left[a_{k}^{\pdag}\,u_{k}^{\pdag}(\eta) + a_{-k}^\dag v_{-k}^{\pdag}(\eta) \right] e^{i \bm{k}\cdot \bm{x}},
\end{eqnarray}
where the operators satisfy $[a_k^{\pdag}, a_q^\dag]_\mp = (2\pi)^3 \delta^{(3)}(\bm{k} - \bm{q})$. The object of interest is the mode function $u_k(\eta)$, which is a complex scalar for $\phi$ and a spinor for $\chi$, with $v_{k} = - i \gamma^2 u_{k}^\star$. Interestingly, in this basis the mode function sastisfy a very simple equation
\begin{equation}\label{eq:EoM_harmonic_oscillator}
{u}_k''(\eta) + \omega_k^2(\eta)\, u_k^{\ptwo}(\eta) = 0,
\end{equation}
and similarly for the left-handed chirality component $x_k$ of the spinor $u_k$ (see however app. \ref{app:conventions_and_fields} for more details). 

All the physics is encoded in the frequency $\omega_k$, which knows about the type of field $\tilde X$, and, together with the boundary conditions, determines the solution for $u_k$.  With the explicit solution is possible to expand the (operator) $\mathcal{H}$ as follows
\al{\label{eq:H}
    \mathcal{H}  
     & = \frac{1}{2}\int\frac{d^3k}{(2\pi)^3} \bigg[ \Omega_k^{\pdag}(\eta)\,(a_{k}^\dag a_{k}^{\pdag} \pm a_{k}^{\pdag} a_{k}^\dag)  \\
     & \hspace{3cm}+ \mathcal{F}_k^{\pdag}(\eta)\, a_{k}^{\pdag} a_{-k}^{\pdag} + h.c. \bigg] .}
This is what we need to compute the average on the proper state to obtain the energy density as in \eqref{eq:energy_density}. Knowing the solution $u_k$ and the initial state, the energy density can be immediately computed. In the bosonic case (spin-0 and spin-1) we have
\be
\Omega_k(\eta) \equiv |u_k'|^2 + \omega_k^2 |u_k^{\ptwo}|^2 , ~~~~~~ \mathcal{F}_k(\eta) \equiv (u_k')^2 + \omega_k^2 u_k^2,
\ee
while for a spin-1/2 spinor we have
\be
\begin{aligned}\label{eq:Omega_F_fermions}
\Omega_k(\eta) &  \equiv  \frac{|x_{k}'|^2 + \omega_k^2 |x_{k}^{\pdag}|^2 - \omega_k^2}{a m} , \\
\mathcal{F}_k(\eta) & \equiv \frac{\xi_k}{k} \bigg[  (x_{k}')^2 + \omega_k^2 x_{k}^2 \bigg].
\end{aligned}
\ee
In the latter expression, $\xi_k$ is a phase satisfying the identities\,\cite{Adshead:2015kza}
$\xi_k^\star \xi_k^{\pdag} = 1$ and $\xi_{-k} = - \xi_k$ \cite{Adshead:2015kza}.
We again refer the reader to app.\,\ref{app:conventions_and_fields} for details and complete derivation of the expressions above.

For the present discussion, all that matters is that, for all cases, far in the past when the modes are deep inside the horizon we can approximate $\omega_k^2 \simeq k^2$, while at late times, when $m \gg H, k/a$, $\omega_k^2 \simeq a^2 m^2$. In these two regimes the frequency is adiabatic, i.e. $\omega_k'/\omega_k^2 \ll 1$, and the concept of particle makes sense. The choice of mode functions defines a specific annihilation operator $a_k$, which in turn defines a vacuum state via $a_k \ket{0_a} = 0$. In Minkowski spacetime, the frequency is time-independent ($\omega_k^2 = k^2 + m^2$) and so is the Hamiltonian, hence, once we choose positive-energy mode functions $u_k \propto e^{-i\omega_k\eta}$, these define the Minkowski vacuum and minimize the energy for all times. In an expanding cosmological background, the situation is more involved:
mode functions that define a reasonable notion of vacuum at early times will, in general, not coincide with the mode functions that define the vacuum at late times. Suppose the latter is defined by mode functions $w_k(\eta)$ associated with annihilation operators $b_k$. A good decomposition of the field $\phi$  thus
\be
\phi(x) = \int\frac{d^3k}{(2\pi)^3} \left[b_{k}^{\pdag}\,w_{k}^{\pdag}(\eta) + b_{-k}^\dag w_{-k}^{\star\pdag}(\eta) \right] e^{i \bm{k}\cdot \bm{x}},
\ee
with the $b_k$ operator (and hence the mode functions $w_k$) identifying a late-time vacuum $b_k \ket{0_b} = 0$. In general, due to the time dependence of the background evolution, one finds $\ket{0_a} \neq \ket{0_b}$, showing that early and late time observers to not agree on the notion of vacuum. 
Since both $\{u_k, u_k^\star\}$ and $\{w_k, w_k^\star\}$ are solutions of eq.\,\eqref{eq:EoM_harmonic_oscillator}, it must be possible to express one in terms of the other via a Bogoliubov transformation
\be\label{eq:Bogoliubov_mode_func}
w_k^{\pstar} = \alpha_k^{\pstar} u_k^{\pstar} - \beta_k^{\pstar} \xi_k^{\pstar}\, u_{-k}^\star.
\ee 
This, in turn, results in a Bogoliubov transformation between $a_k$ and $b_k$ operators:
\be\label{eq:Bogoliubov}
\begin{aligned}
    a_k^{\pdag} & = \alpha_k^{\pdag} b_k^{\pdag} + \beta_k^{\star\pdag}\, \xi_k^{\star\pdag}\,  b_{-k}^\dagger, \\
    a_{-k}^\dagger & = \alpha_k^{\star\pdag} \, b_{-k}^\dagger + \beta_k^{\pdag}\,\xi_{-k}^{\pdag}\, b_k^{\pdag}  .
\end{aligned}
\ee
Due to quantization, we have the constraint $|\alpha_k|^2 \mp |\beta_k|^2 = 1$. For bosons, the phase $\xi_k$ is even under $\bm{k} \to - \bm{k}$, as is $\beta_k$, so it can be reabsorbed in the latter parameter. For fermions, this phase is instead instrumental in guaranteeing that the Bogoliubov transformation is not singular.\,\footnote{More precisely, without the property $\xi_{-k} = - \xi_k$, we would obtain that $b_k$ and $b_k^\dag$ would always be proportional to $(|\alpha_k|^2 - |\beta_k|^2)^{-1}$, independently from the statistics. While for bosons this is always equal to one and the transformation is always invertible, for fermions it may vanish, making the Bogoliubov transformation singular. The property $\xi_{-k} = -\xi_k$ guarantees that, for fermions, $b_k^{\pdag}$ and $b_k^\dag$ are proportional to $(|\alpha_k|^2 + |\beta_k|^2)^{-1} = 1$, making the transformation always invertible.} 

In a cosmological background, it is typically not possible to find solutions that define the vacuum minimizing the energy for all times. However, the fact that the frequency varies adiabatically for early and late times allows us to define an adiabatic vacuum using as mode function the positive-energy solution computed via a WKB approximation. At late times we can thus take
\be\label{eq:WKB}
w_k = \frac{1}{\sqrt{2 \omega_k(\eta)}} e^{- i\int^\eta d\eta' \omega_k(\eta')}, 
\ee
while the mode function $u_k$ will be found solving eq.\,\eqref{eq:EoM_harmonic_oscillator} and imposing appropriate initial conditions. Notice that, in terms of the adiabatic mode functions\,\eqref{eq:WKB}, the late-time Hamiltonian reads
\be\label{eq:H_late_times}
\mathcal{H}_{\rm late} = \frac{1}{2} \int \frac{d^3k}{(2\pi)^3} \omega_k^{\pdag}(\eta) \bigg[ b_k^\dag b_k^{\pdag} \pm b_k^{\pdag} b_k^\dag \bigg] + \mathcal{O}\left(\frac{\omega_k'}{\omega_k^2}\right),
\ee
i.e. it is diagonal apart from terms that are suppressed in the adiabatic limit. Again, the upper (lower) sign applies to bosons (fermions).  

Up to this point the discussion has been completely in terms of operators, but in order to compute observables, a state must be specified as well. We work in Heisenberg picture and fix the state to be $\ket{\psi}$ (not necessarily the vacuum) at initial times. 
The number of gravitationally produced particles, as measured by a late time observer that uses the $b_k$ basis (us), can be computed in two different, but equivalent ways. One can, in terms of the mode function $u_k$, equate
the expectation value of eq.\,\eqref{eq:H} with the one of eq.\,\eqref{eq:H_late_times}, as both expressions for the Hamiltonian should represent the same physics. 
From the matching condition
\be
\lim_{\eta\to \infty}\langle \psi|\mathcal{H}|\psi\rangle=\langle\psi| \mathcal{H}_{\rm late}|\psi\rangle,
\ee
we can derive an expression for the final number of particles, by reading out the vacuum expectation value of $b^\dag b$ on $|\psi\rangle$. Defining $\langle N_k^f \rangle \equiv \bra{\psi} b_k^\dag b_k^{\pdag} \ket{\psi}$, we have
\al{\label{eq:N_1}
    \langle N_k^f\rangle  
    & = \frac{1}{2\omega_k} \bigg[\Omega_k^{\ptwo}(\eta)\, \left(2 \langle N_k^\text{in}\rangle \pm V \right)\\
    & \hspace{3cm} + \mathcal{F}_k^{\ptwo}(\eta)\, \langle C_k^\text{in} \rangle + c.c. \bigg] \mp \frac{V}{2}\,.
}
In the above equation we have some important quantities for our discussion: $i)$ $\langle N_k^\text{in}\rangle \equiv \bra{\psi} a_k^\dag a_k^{\pdag} \ket{\psi}$ is the number of particles of comoving momentum $k$ in the initial state; ii) $\langle C_k^\text{in} \rangle \equiv \bra{\psi} a_k a_{-k} \ket{\psi}$. The appearance of comoving volume factors $V$ originates from the use the commutation relations yielding $(2\pi)^3 \delta^{(3)}(0) \equiv V$. The upper (lower) sign applies to bosons (fermions). Alternatively, $\langle N_k^f \rangle$ can be obtained directly in terms of the Bogoliubov transformation of eq.\,\eqref{eq:Bogoliubov} via
\al{\label{eq:N_2}
    \langle N_k^f \rangle 
    & = V  |\beta_k^{\ptwo}|^2 + \left(1 \pm 2 |\beta_k^{\ptwo}|^2\right) \langle N_k^\text{in} \rangle \\
    & \hspace{3cm} - \left( \alpha_k^\star \,\beta_k^\star\,\xi_k^\star\, \langle C_k^\text{in} \rangle+c.c. \right),
}
where we have used $|\alpha_k|^2 \mp |\beta_k|^2 =1$. Given our definition of the Bogoliubov transformation and of the adiabatic vacuum, eqs.\,\eqref{eq:N_1} and\,\eqref{eq:N_2} are equivalent. 

Given the stochastic and isotropic nature of GPP, a natural definition for the number density emerges as
\be\label{eq:na3}
n\,a^3 = \int \frac{d^3k}{(2\pi)^3} \frac{\langle N_k^f \rangle}{V} = \int \frac{dk}{k} \left(\frac{k^3}{2\pi^2} n_k \right),
\ee
where in the last step we have defined $n_k = \langle N_k^f\rangle/V$ and $a$ is the scale factor. If we now define the initial (comoving) number density $\langle n_k^\text{in}\rangle \equiv \langle N_k^\text{in}\rangle/V$ and $\langle c_k^\text{in} \rangle \equiv \langle C_k^\text{in} \rangle/V$, we can write 
\al{\label{eq:nk_1}
    n_k  
    & = \frac{1}{2\omega_k} \bigg[\Omega_k(\eta) \left(2 \langle n_k^\text{in}\rangle \pm 1 \right)\\
    & \hspace{3cm} + \mathcal{F}_k(\eta) \langle c_k^\text{in} \rangle + c.c. \bigg] \mp \frac{1}{2} ,
}
or, equivalently,
\al{\label{eq:nk_2}
    n_k 
    & =  |\beta_k^{\ptwo}|^2 + \left(1 \pm 2 |\beta_k^{\ptwo}|^2\right)  \langle n_k^\text{in}\rangle\\
    & \hspace{3cm} -  \left( \alpha_k^\star\, \beta_k^\star\,\xi_k^\star\, \langle c_k^\text{in} \rangle +c.c \right).
}
As a crosscheck, we observe that, when $\ket{\psi}$ is the Bunch--Davies vacuum, we have $\langle n_k^\text{in} \rangle = \langle c_k^\text{in} \rangle = 0$ and $2 \pi^2 n_k/k^3 = |\beta_k|^2 = (\Omega_k - \omega_k)/(2\omega_k)$.\,\footnote{In Minkowski spacetime, $\Omega_k = \omega_k$ and there is no CGGP as expected.} Actually, for the Bunch-Davies vacuum we also have that $\mathcal{F}_k(\eta\to -\infty)=0$. This is the standard result used in the literature\,\cite{Kolb:2023ydq}.

When we allow for a more generic initial state, two effects emerge: (i) a ``stimulated emission'' or ``Pauli blocking'' term proportional to $\langle n_k^\text{in}\rangle$ (this nomenclature was introduced already in ref.\,\cite{Parker:1969au}) and (ii) terms proportional to the matrix elements $\bra{\psi} a_k a_{-k} \ket{\psi}$ and $\bra{\psi} a_k^\dagger a_{-k}^\dagger \ket{\psi}$, which can be non-vanishing only if the initial state $\ket{\psi}$ is a linear combination of multiparticle states of the form
\be\label{eq:psi_state}
\ket{\psi} = \sum_n \int \frac{d^3k_1}{(2\pi)^3} \dots \frac{d^3k_n}{(2\pi)^3} \psi_n(\bm k_1, \dots, \bm k_n) \ket{\bm k_1, \dots, \bm k_n} ,
\ee
where the functions $\psi_n(\bm k_1, \dots, \bm k_n)$ must be chosen in such a way that $\braket{\psi|\psi} = 1$ and have the appropriate symmetry properties to satisfy the statistics of the field. They must also be chosen to ensure that the expectation value of the energy-momentum tensor is finite. The conclusion is clear: in order to compute the final number of gravitationally produced particles, it is sufficient to solve the EoM\,\eqref{eq:EoM_harmonic_oscillator} (and the analogous ones for fermions) and use eq.\,\eqref{eq:nk_1}. Alternatively, we can compute the Bogoliubov coefficients $\alpha_k$ and $\beta_k$ and use eq.\,\eqref{eq:nk_2}. The two approaches are completely equivalent.

Before concluding the discussion, we make two final comments. First, we observe that states like the one in eq.\,\eqref{eq:psi_state} i.e. superpositions of multiparticles states, are not so exotic as they may seem. Indeed, we can list at least two well-known cases in which this happens: (i) squeezed states like those obtained via a Bogoliubov transformation (even those obtained via GPP)\,\cite{Ford:2021syk} and (ii) coherent states. The first case (together with a case in which $\langle C_k^\text{in} \rangle = 0$) will be considered in sec.\,\ref{sec:results}, while we argue in app.\,\ref{app:coherent} that the coherent states have no impact in the relic abundace. Second, we observe that the computation of the spectrum from the expressions in eqs.\,\eqref{eq:nk_1}-\eqref{eq:nk_2} carries a redundancy. In other words, one is allowed to choose how to read the non-trivial initial conditions, depending to which term in $n_k$ carries the information about them. There are basically two different interpretations: 
\begin{itemize}
    \item The state $\ket{\psi}$ carries all the information about the initial conditions and we take Bunch--Davies initial conditions for the mode functions, namely $u_k(\eta_i)= e^{-i\omega_k(\eta_i)\eta_i}/\sqrt{2\omega_k(\eta_i)}$ and $u_k'(\eta_i)=-i\omega_k(\eta_i) u_k(\eta_i)$, with $\eta_i$ the initial conformal time. In this way, $\Omega_k(\eta_i)=\omega_k(\eta_i)$ and $\mathcal{F}_k(\eta_i)=\beta_k(\eta_i)=0$, whereas we take $n_k(\eta_i)=\braket{n_k^\text{in}}$ and $\braket{c_k^\text{in}}\neq 0$, such that eqs.\,\eqref{eq:nk_1}-\eqref{eq:nk_2} are automatically satisfied at initial times;

    \item The state is taken to be the vacuum $\ket{\psi}=\ket{0_\text{in}}$, while the mode functions $u_k$ carry all the information about the initial state. As a consequence, $\langle n_k^\text{in} \rangle = 0$ and $\langle c_k^\text{in} \rangle = 0$ as if we were considering a Bunch--Davies initial state, but we choose initial conditions $u_k(\eta_i)$ and $u_k'(\eta_i)$ in such a way that $\Omega_k(\eta_i) = \omega_k(\eta_i) (\pm \,2n_k(\eta_i) + 1)$ or, equivalently, $|\beta_k(\eta_i)|^2=n_k(\eta_i)$. This again guarantees that eqs.\,\eqref{eq:nk_1}-\eqref{eq:nk_2} are satisfied at initial times. 
\end{itemize}
In practical computations, it might be more convenient to use one of the interpretations above rather than the other. We have checked for several cases, including the ones discussed in sec.\,\ref{sec:thermal}, that we obtain the same results with both approaches. In studies involving more observables other than the spectrum, possibly this redundancy is broken and one might have to choose only one of these interpretations, thus giving more information on the structure of the initial conditions of the relics. We leave such analysis for future work.

\subsection{Power spectrum}\label{sec:power_spectrum}
An alternative way of performing the computation employs the power spectrum of the field, when sizable production happens during inflation, as it is for minimally coupled scalars. For simplicity of notation, we will from now on drop the tilde from Weyl-rescaled fields. For the case of a bosonic $X$, we have
\be\label{eq:power_spectrum}
\langle  X_k(\eta)  X_q(\eta) \rangle  = (2\pi)^3 \delta^{(3)}(\bm{k}+\bm{q}) \frac{2\pi^2}{k^3} \mathcal{P}_{ X}(k,\eta),
\ee
where the average is taken over the state $\ket{\psi}$. Our starting point is again the Hamiltonian, written using the field basis in which the EoM is of the harmonic oscillator type.  Taking the expectation value between $\ket{\psi}$ states, the result can be expressed using the power spectrum defined in eq.\,\eqref{eq:power_spectrum} as
\be\label{eq:H_power_spectrum}
\langle \mathcal{H} \rangle = \frac{V}{2}\int \frac{dk}{k} \left[ \mathcal{P}_{ X'} + \omega_k^2\, \mathcal{P}_{ X} \right],
\ee
where $V$ is, again, the comoving volume over which we are integrating. This result applies to bosons, while we defer the discussion of the fermionic case to app.\,\ref{app:spin1/2}. From this expression it is clear that the power spectrum of the field contains the same information contained in the number of particles, with the advantage that, while the number of particles is well defined only when the evolution is adiabatic, the power spectrum is always well defined. We can make the comparison more precise focusing on the scalar $\phi$, comparing eq.\,\eqref{eq:H_power_spectrum} with the expectation value of eq.\,\eqref{eq:H_late_times}, from which we obtain
\be
n_k = \frac{1}{2} \bigg[ \frac{2\pi^2}{k^3} \frac{1}{\omega_k} \bigg( \mathcal{P}_{\phi'} + \omega_k^2\, \mathcal{P}_\phi \bigg)-1\bigg].
\ee
Again, we stress that this expression can be interpreted as the particle spectrum only at late times, when the evolution is adiabatic and the concept of particle well defined. In this regime, the field evolution is virialized ($\mathcal{P}_{\phi'} \sim \omega_k^2 \mathcal{P}_\phi$) and we have $(k^3/2\pi^2) n_k \simeq \omega_k \mathcal{P}_\phi$. In sec.\,\ref{sec:results}, we will present plots for $(k^3/2\pi^2) n_k$ which is a probe for the power spectrum at late times. The power spectrum also encodes information about the initial conditions. To see this, we observe that the average in eq.\,\eqref{eq:power_spectrum} can be made explicit using the field decomposition of eq.\,\eqref{eq:quantum_field}. The matrix elements of the bilinears of annihilation and creation operators that appear are simplified using momentum conservation: 
\be
\begin{aligned}
    \langle a_k^\dag a_q \rangle & = \frac{\langle a_k^\dag a_k^{\pdag} \rangle}{V} \,(2\pi)^3 \delta^{(3)}(\bm{k} - \bm{q}), \\
    \langle a_k a_q \rangle & = \frac{\langle a_k a_{-k} \rangle}{V} \, (2\pi)^3 \delta^{(3)}(\bm{k} + \bm{q}), 
\end{aligned}
\ee
from which we obtain, for any bosonic field, 
\be\label{eq:power_spectrum_NBD}
\begin{aligned}
    \frac{2\pi^2}{k^3} \mathcal{P}_{ \phi}(k,\eta)&  = |u_k^{\pdag}(\eta)|^2 \left( 2\langle n_k^\text{in} \rangle  + 1 \right) \\
    & \quad + u_k^{\pdag}(\eta)^2\,  \braket{c_k^\text{in}}  + c.c. .
\end{aligned}
\ee
We thus conclude that the power spectrum contains, at the same time, information about the initial conditions (through $\langle n_k^\text{in} \rangle$ and $\langle c_k^\text{in} \rangle$) and the subsequent evolution (through $u_k(\eta)$). Once these information are known, $\mathcal{P}_X(k,\eta)$ can be computed at any time. Since we are concerned with imposing appropriate initial conditions, it is convenient to follow the standard procedure and introduce a transfer function $T(k, \eta)$ such that
\be\label{eq:power_spectrum_transfer_func}
\mathcal{P}_X(k, \eta) = \mathcal{P}_X(k)_\text{exit}\, T(k, \eta),
\ee
where $\mathcal{P}_X(k)_\text{exit}$ is the power spectrum at the moment of horizon exit during inflation.
With respect to the standard case there is a complication. When the term proportional to $\langle c_k^\text{in} \rangle$ is absent in eq.\,\eqref{eq:power_spectrum_NBD}, the transfer function can be simply computed as $T(k, \eta) = |u_k(\eta)|^2/|u_k(\eta_i)|^2$. When $\langle c_k^\text{in} \rangle \neq 0$, care must be taken, since the terms $|u_k|^2$ and $u_k^2$ in eq.\,\eqref{eq:power_spectrum_NBD} agree on the overall dependence on $a$, but evolve with different phases. As we are going to see in sec.\,\ref{sec:results}, this will cause an oscillatory behavior in the final results but does not change the overall dependence on the scale factor. 
\begin{figure}[t]
\centering
\includegraphics[width=0.48\textwidth]{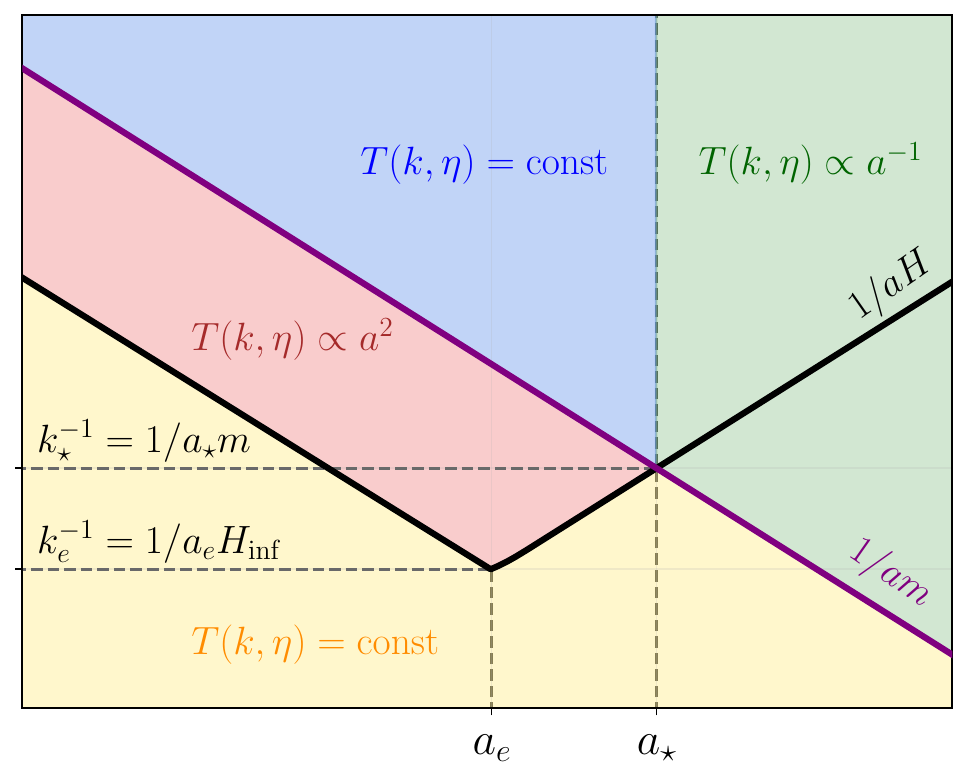}
\caption{\label{fig:power_spectrum_standard} Sketch showing the behavior of the transfer function of a vector field, eq.\,\eqref{eq:transfer_function_vector_text} for instantaneous reheating. We show the comoving horizon $1/aH$ (black) and the comoving Compton wavelenght $1/am$ (purple). The scale factor $a_e$ denotes the end of inflation and $a_\star$ the moment in which $m=H$ and the evolution becomes adiabatic. We take $H_\text{inf}$ to be the Hubble parameter during inflation.}
\end{figure}

\subsubsection{Peak in the power spectrum: the case of vector DM}
Before moving to the explicit examples in section \ref{sec:results}, we would like to comment on the case of a massive vector, with notation as in app. \ref{app:spin-1}. This case is a reference for all models that develop a peak in the power spectrum at some comoving momentum $k_{\rm peak}$. This happens for all fields, with the exception of the minimally coupled scalar. The appearance of the peak can be understood inspecting the transfer functions $T(k, \eta)$, which are reported in app.\,\ref{app:conventions_and_fields} for various fields. Here we discuss in detail the case of spin-1, but similar reasoning can be applied to the other cases. This also allows us to reconnect with the famous case of dark photon DM production from inflationary perturbations \cite{Graham:2015rva}.

Suppose we are given the initial power spectrum at horizon exit during inflation. As in the appendix, when discussing analytic results we approximate inflation as a pure de Sitter phase of expansion (with Hubble parameter $H_\text{inf}$) and take reheating to be instantaneous. The transfer function for the Weyl-rescaled longitudinal component of a vector field, called $u$ in app.\,\ref{app:conventions_and_fields}, is (see eq.\,\eqref{eq:transfer_function_vector})
\be\label{eq:transfer_function_vector_text}
T(k, \eta) \propto \left\{
\begin{array}{ccc}
    \text{const} & ~~~ & k \gg aH,\,am   \\
    a^2 &  ~~~ & aH \gg k \gg am \\
    \text{const} & ~~~ & aH \gg am \gg k \\
    a^{-1} & ~~~ & am \gg k,\,aH
\end{array}\right.
\ee
where we have maintained only the overall $a$ scaling, averaging over oscillations. We find more useful to parametrize the transfer function in terms of the scale factor, rather than conformal time. With the results above, we can follow the evolution of different modes $k$ using fig.\,\ref{fig:power_spectrum_standard} as a guide, identifying three relevant regimes of comoving momenta. For modes $k < k_\star$, the modes enter the red region at horizon exit ($a = k/H_\text{inf}$), they evolve as $a^2$ until the moment in which they enter the blue region ($a = k/m$), at which point they remain constant until crossing into the green region, in which they start decreasing as $a^{-1}$. We thus have
\be
T(k, \eta) = \left(\frac{k/m}{k/H_\text{inf}}\right)^2 \frac{a_\star}{a} = \frac{H_\text{inf}^2}{m^2} \frac{a_\star}{a}, ~~~~~~~~~ k < k_\star.
\ee
For modes $k_\star < k < k_e$, the modes again exit the horizon into the red region when $a = k/H_\text{inf}$. They then evolve as $a^2$ until they enter again the horizon during radiation domination (when $a = a_e^2 H_\text{inf}/k$), they remain constant while they are inside the horizon and then enter the green region at $a = k/m$, at which point they start to evolve as $a^{-1}$. Overall, the transfer function for these modes results in
\be
T(k, \eta) = \left(\frac{a_e^2 H_\text{inf}/k}{k/H_\text{inf}}\right)^2 \frac{k/m}{a} = \frac{(a_e H_\text{inf})^4}{k^3\, m\, a} , ~~~~~~ k_\star < k < k_e.
\ee
Finally, for modes $k > k_e$ the evolution is always adiabatic and no GPP happens.

Putting everything together, we have that the transfer function is initially constant and then decreases as $k^{-3}$. This behavior contributes to the formation of the peak previously mentioned. To further clarify this point, we consider the usual case of Bunch--Davies initial conditions, i.e. we take $u_k(\eta_i) \to e^{-ik\eta_i}/\sqrt{2k}$ (and $\braket{N_k^\text{in}}=\braket{C_k^\text{in}}=0$) in the left part of the yellow region of fig.\,\ref{fig:power_spectrum_standard}. For this initial condition, the power spectrum at horizon exit is simply
\be\label{eq:Pk_initial_BunchDavies}
\mathcal{P}_u(k)_\text{exit} = \frac{k^2}{4\pi^2} ~~~~~~~ \text{(Bunch--Davies)}.
\ee
Combining this with the transfer function computed above, we obtain a power spectrum that grows as $k^2$ for $k < k_\star$ and decreases as $k^{-1}$ for $k> k_\star$, i.e. the spectrum is peaked as anticipated. For future convenience, we use the results obtained so far to estimate analytically the late-time DM energy density for Bunch--Davies initial conditions. This can be done using eq.\,\eqref{eq:na3} and estimating $(k^3/2\pi^2) n_k \simeq \omega_k \mathcal{P}_\phi/2$ using the power spectrum just computed and $\omega_k \simeq a\,m$. Integrating over all momenta we obtain 
\be\label{eq:na3_vector_standard}
n a^3 \simeq \frac{3(a_e H_\text{inf})^3}{8\pi^2} \sqrt{\frac{H_\text{inf}}{m}},
\ee
from which the following final abundance can be derived:
\al{\label{eq:Omega_vector_standard}
&\frac{\Omega_\text{vector}}{\Omega_\text{CDM}} \simeq\\
&\hspace{0.7cm}\left(\frac{T_\text{RH}}{10^9\,\text{GeV}}\right)\left(\frac{H_\text{inf}}{10^{12}\,\text{GeV}}\right)^2 \frac{m}{H_\text{inf}} \frac{na^3/(a_e H_\text{inf})^3}{10^{-5}},
}
with $\Omega_\text{CDM}\simeq 0.26$ the present observed DM abundance.
We observe that, comparing eqs. \eqref{eq:na3_vector_standard} and \eqref{eq:Omega_vector_standard}, we reproduce the well-known behavior $\Omega \propto \sqrt{m}$\,\cite{Graham:2015rva,Ahmed:2020fhc,Kolb:2020fwh,Redi:2022zkt}. 
Although we have shown explicitly only the case of instantaneous reheating, a similar analysis can be performed for finite reheating and the spectrum is similarly peaked at $k_\star$. In what follows, it will be useful to have approximate expressions for $k_\star$:
\al{\label{eq:kstar}
k_\star & \simeq a_eH_\text{inf}\left(\frac{m}{H_\text{inf}}\right)^{1/3}, & H=m~\text{at~reheating},\\
k_\star & \simeq a_eH_\text{inf}\left(\frac{m}{H_\text{inf}}\right)^{1/2}\left(\frac{H_\text{inf}}{H_\text{RH}}\right)^{1/6}, & H=m~\text{at~radiation},
}
where $H_\text{RH}=\sqrt{\pi^2g_*(T_\text{RH})T_\text{RH}^4/(90M_P^2)}$ is the Hubble at the end of reheating, with $M_P$ the reduced Planck mass, and $g_*(T)$ the number of relativistic degrees of freedom at temperature $T$.

Of course, while the conclusion that the power spectrum at late times is peaked at $k_\star$ relies on the usual Bunch--Davies initial conditions in eq.\,\eqref{eq:Pk_initial_BunchDavies}, the transfer function $T(k,\eta)$ we have computed is, except for oscillations, applicable to any initial condition. In particular, given a more generic initial power spectrum, different from eq.\,\eqref{eq:Pk_initial_BunchDavies}, its convolution with the transfer function does not necessarily produce a peak at $k_\star$, or may even produce multiple peaks. In fact, both the explicit examples considered in sec.\,\ref{sec:results} will have peaked power spectra, one for which $k_\text{peak}=k_\star$ as before, and another where a second peak will appear, located at another comoving momentum. As we are going to discuss in detail, the peaked behavior will be instrumental in determining the experimental limits on the scenario considered.


\section{Results}\label{sec:results}

In this section we present our results by illustrating the impact of initial conditions on the production of relics via GPP. We consider two different cases: in the first, we will assume a initial thermal distribution without specifying its origin; in the second, we will instead assume a two-stage inflation model with an extra phase of radiation domination between the quasi de Sitter evolution. We will analyze in detail both possibilities, stressing the new features brought by the new initial conditions.

We start with a general comment about fields for which the only break of conformality is due to their mass (conformally coupled scalar with $\xi = 1/6$, spin 1/2 spinors, transverse polarizations of vector fields). For these cases, $\omega_k \simeq k$ until the mode crosses the comoving Compton wavelength and the mass term in $\omega_k$ becomes important (see app.\,\ref{app:conventions_and_fields}), so that the mode functions that appear in eqs.\,\eqref{eq:nk_1}-\eqref{eq:nk_2} are the Bunch--Davies ones to a very good degree of approximation. This implies $\Omega_k \simeq \omega_k$ and  $\mathcal{F}_k \simeq 0$, so that there is essentially no GPP in this regime and $n_k \simeq \langle n_k^\text{in}\rangle$ until Compton crossing. On general ground, we expect physically relevant modifications to $\langle n_k^\text{in}\rangle$ to happen for relatively small $k$. On the other hand, after Compton crossing, for the cases under consideration $|\beta_k|^2$ peaks at large $k_\star$, corresponding to large masses, in such a way that the product $|\beta_k|^2 \langle n_k^\text{in}\rangle$ appearing in eq.\,\eqref{eq:nk_2} is much smaller than the pure GPP contribution $|\beta_k|^2$. We thus expect these ``almost-conformal'' field to be only slightly affected by non-Bunch-Davies initial conditions. This will be the case in the thermal example considered in sec.\,\ref{sec:thermal}, for which we numerically checked that a thermal initial distribution have almost no effect on the final abundance of conformally coupled scalars. In the example discussed in sec.\,\ref{sec:pre_inflation}, on the other hand, the two-stage inflationary scenario considered is such that no GPP at all happens before Compton crossing, making our claim exact.
We also do not focus on minimally coupled scalars, since it is known that they are excluded by isocurvature constraints\,\cite{Chung:2004nh,Chung:2011xd,Garcia:2023qab} (see app.\,\ref{app:isocurvature}) and a non-vanishing initial population does not change this conclusion. Different is the situation of the longitudinal components of massive vectors: they are, in general, not affected by isocurvature perturbations and their GPP starts well before the mass becomes important, so that they are most affected by non Bunch--Davies initial conditions.

For all our results, we consider a period of inflation, described for simplicity by the single-field quadratic (chaotic) model, followed by a phase of reheating propelled by inflaton decays and by the usual period of radiation domination.\footnote{Even though this quadratic inflationary model is excluded by current experimental data\,\cite{Planck:2018jri}, the relics produced from GPP are rather insensitive to the precise dynamics of inflation for masses $m<H_\text{inf}$, which is the regime considered here.}  In what follows, $a_e$, $H_\text{inf}$ ($T_\text{RH}$, $H_\text{RH}$) denote the scale factor and Hubble parameter (temperature and Hubble) at the end of inflation (reheating).

\subsection{Thermal distribution}\label{sec:thermal}

We first assume the non Bunch--Davies initial condition to be a thermal distribution. More precisely, we take
\be\label{eq:thermal_distribution}
\braket{n_k^\text{in}} = \frac{1}{e^{k/T}-1}, \quad \braket{c_k^\text{in}} = 0,
\ee
at some initial time $a_i$. These relations follow computing the expectation values of an operator $\mathcal{O}$ as $\text{tr}(\rho \mathcal{O})$ with thermal density matrix
\be
\rho = \prod_k \left(1-e^{-k/T}\right) \sum_{N_k = 0}^\infty e^{- N_k k/T} \ket{N_k}\bra{N_k},
\ee
where $\ket{N_k}$ is a state with $N_k$ quanta with momentum $k$. The parameter $T$ is the comoving temperature, related as usual to the physical temperature of the pre-inflationary thermal bath by $T_\text{phys} =  T/a$. To be independent of the convention of the scale factor, we present our result in terms of $T/(a_eH_\text{inf})$, which is a physical dimensionless quantity.
Here we remain agnostic about the origin of eq.\,\eqref{eq:thermal_distribution}, but one can imagine that it could have been generated from the dynamics of a dark sector that produced the relic through a thermal mechanism. For our computation to be valid, we need to assume that the interactions that generated the thermal bath are switched off at the time we impose initial conditions, so that we can safely use the free EoMs for the mode functions.

\begin{figure}[t]
\centering
\includegraphics[width=0.48\textwidth]{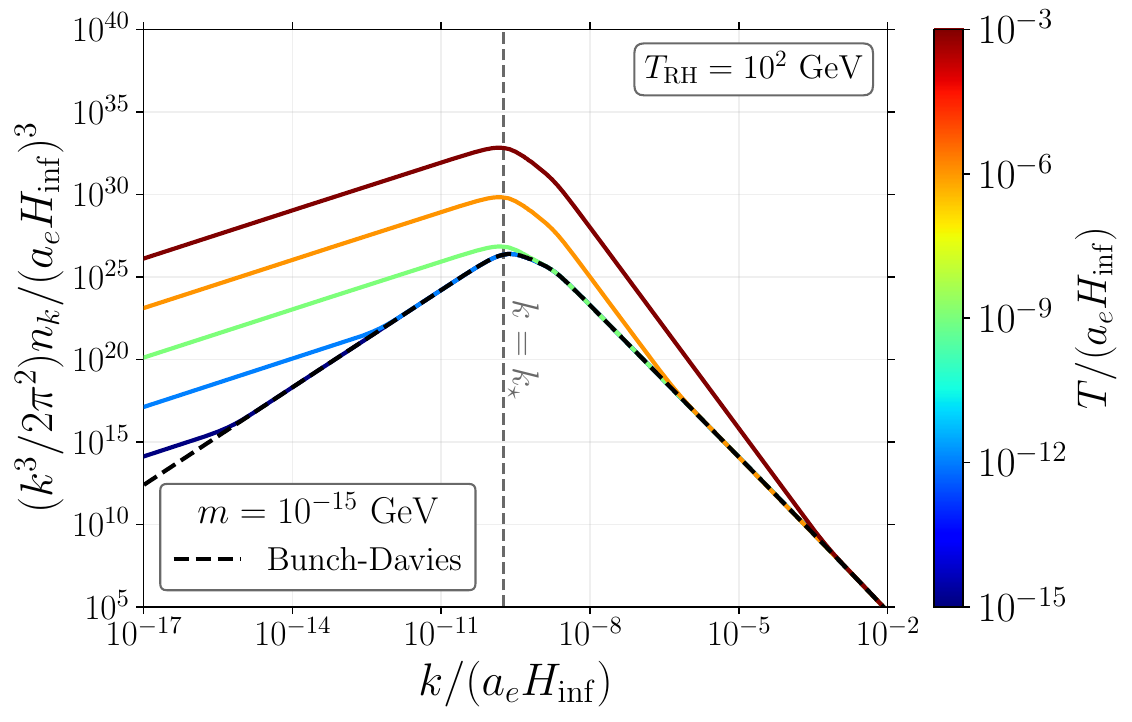}
\caption{Comoving momentum spectrum for a spin-1 relic with a mass $m=10^{-15}$ GeV and reheating temperature $T_\text{RH}=10^{2}$ GeV. Dashed black denotes results obtained using Bunch--Davies initial conditions, while colors assume different values for the initial temperature $T$ according to the color code.}\label{fig:Spectrum_thermal}
\end{figure}

\begin{figure}[t]
\centering
\includegraphics[width=0.48\textwidth]{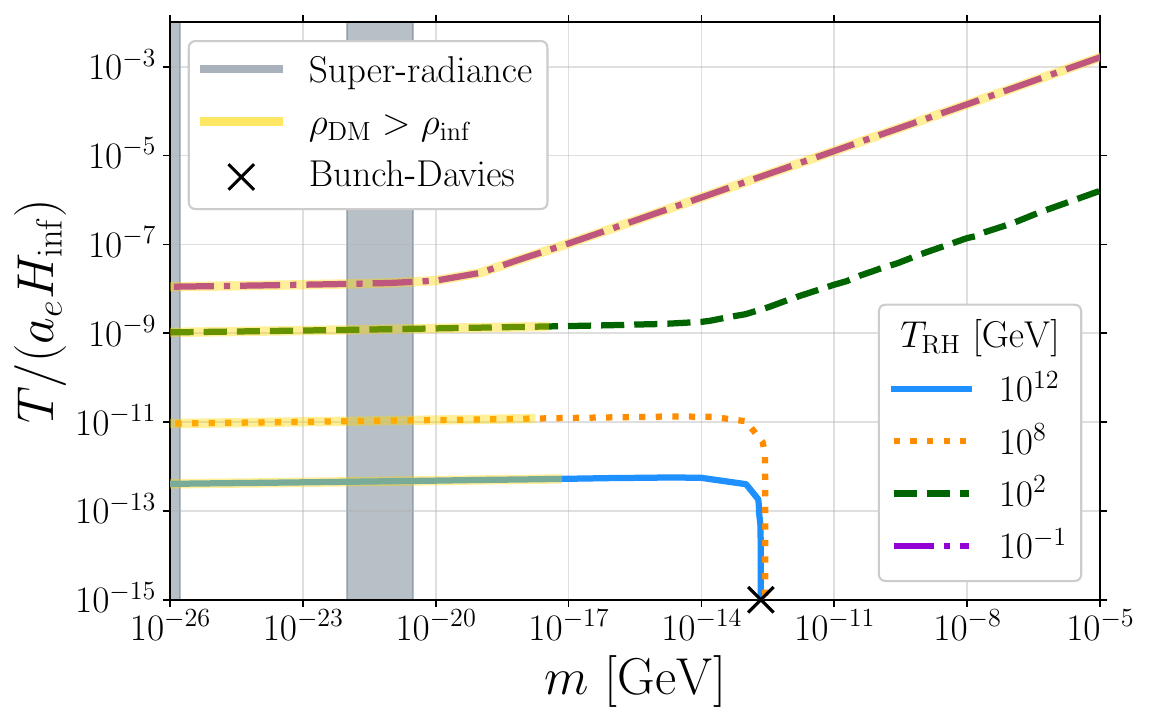}
\caption{Abundance levels for spin-1 DM in the scenario with an initial thermal distribution. We show in the $m-T$ parameter space the curves representing the total DM abundance today. Colors denote different reheating temperatures, namely $T_\text{RH} = 10^{12}$ GeV (solid blue), $T_\text{RH} = 10^{8}$ GeV (dotted orange), $T_\text{RH} = 10^{2}$ GeV (dashed green) and $T_\text{RH} = 10^{-1}$ GeV (dashed-dotted purple). We fix $H_\text{inf}=10^{13}$ GeV. The gray regions are excluded by super-radiance limits, while the yellow bands by eq.\,\eqref{eq:thermal_bound_density}, see text for details. The black cross denotes the mass $m\simeq 2\times 10^{-13}$ GeV for which Bunch--Davies initial conditions can produce all the DM.}\label{fig:Oh2_thermal}
\end{figure}

In fig.\,\ref{fig:Spectrum_thermal} we present our results for the normalized momentum spectrum $(k^3/2\pi^2) n_k/(a_eH_\text{inf})^3$ of the longitudinal component of a vector particle, for concreteness choosing $m=10^{-15}$ GeV and $T_\text{RH}=10^{2}$ GeV. The dashed line shows the spectrum calculated with Bunch--Davies initial conditions (i.e. no initial particle population), while the colored lines show the spectrum obtained with different normalized initial temperature $T/(a_eH_\text{inf})$. From the plot we see that, although the slopes change with respect to the usual Bunch--Davies case, the thermal distribution preserves most of the features of the spectrum. In particular, the peak is still located at $k_\star$, as indicated by the vertical dashed gray line. This can be understood from the fact that the impact of the initial thermal distribution is only important when $k\ll T$, implying that $\braket{n_k^\text{in}}\simeq T/k$, which is why the slopes of $n_k$ change. Since $(k^3/2\pi^2)n_k|_\text{Bunch-Davies}\sim k^2$ for $k<k_\star$ and decreases afterwards (see discussion after eq.\,\eqref{eq:Pk_initial_BunchDavies}), the peak will remain at $k_\star$, as seen from fig.\,\ref{fig:Spectrum_thermal}.  The thermal distribution stops being relevant for masses that satisfy $k_\star\gtrsim T$. Nevertheless, we see that the initial thermal distribution has an impact for $k \lesssim T$ as long as the GPP peak falls in this region, i,.e. $k_\star \lesssim T$. If this is not the case, the impact of thermal initial conditions is small and resembles the one of ``almost conformal'' fields discussed above.

Moving on to fig.\,\ref{fig:Oh2_thermal}, we show the spin-1 abundance curves corresponding to the total DM today in the parameter space of mass and initial temperature. The different curves correspond to different reheating temperatures $T_\text{RH}$ for a fixed $H_\text{inf}=10^{13}$ GeV. We distinguish between two regimes: high and low reheating temperatures. In the former case ($T_\text{RH}=10^{12}$ GeV, solid blue, and $T_\text{RH}=10^{8}$ GeV, dotted orange) the curve remains horizontal for low masses and then suddenly drops at a certain mass value. This drop takes place precisely at the mass that would correspond to the total DM in the Bunch--Davies scenario, indicated by the cross $\times$, above which there is over-production of DM. The approximate constant behavior follows from
\begin{align}
\begin{aligned}\label{eq:Omega_thermal_estimate}
\Omega
& \propto m~\int\frac{\dd k}{k}\frac{k^3}{2\pi^2}n_k\\
& \simeq m \int\frac{\dd k}{k}~\frac{k^3|\beta_k|^2}{2\pi^2}\braket{n_k^\text{in}}\\
&\propto \Omega|_\text{Bunch-Davies} \frac{T}{k_\star},
\end{aligned}
\end{align}
where $\Omega|_\text{Bunch-Davies}$ is the abundance for Bunch--Davies initial conditions. In the equation above, we have just considered the contribution from the ``stimulated production" term, used that the integral is dominated for $k\simeq k_\star$ and assumed $T\gg k_\star$. For small masses, i.e. when $H=m$ happens during radiation domination, $\Omega|_\text{Bunch-Davies}/k_\star \sim \sqrt{m}/\sqrt{m}=\text{constant}$, see eqs.\,\eqref{eq:na3_vector_standard} and\,\eqref{eq:Omega_vector_standard}, explaining why $\Omega$ is independent of the mass in this regime. In the case of low reheating temperatures (dashed-dotted purple for $T_\text{RH}=10^{-1}$ GeV and dashed green for $T_\text{RH}=10^2$ GeV), the constant behavior for small masses is identical, but after a certain point the curves instead grow with temperature and mass. This is a consequence of the fact that, for such low reheating, the usual GPP with Bunch--Davies initial conditions cannot produce the entirety of the DM, needing thus to rely on the enhancement through the initial conditions to achieve the correct abundance today. Following a similar estimate from eq.\,\eqref{eq:Omega_thermal_estimate} but for masses that cross the threshold $H=m$ during reheating, we find $\Omega \propto \Omega|_\text{Bunch-Davies}T/k_\star \sim T/m^{1/3}$ --see eq.\,\eqref{eq:kstar}--, which correctly describes fig.\,\ref{fig:Oh2_thermal}. Furthermore, since $T/k_\star \sim T_\text{RH}^{2/3}T$, the lower the reheating temperature is, the larger the initial temperature must be to generate the same abundance.

In fig.\,\ref{fig:Oh2_thermal} we also show some constraint on our parameter space. First, the final spectrum $(k^3/2\pi^2)n_k$ must remain integrable in order to avoid a catastrophic production of particles. This is evident from fig.\,\ref{fig:Spectrum_thermal}, which shows that the spectrum is peaked at $k_\star$ and falls to zero for both small and large $k$ and therefore can be safely integrated over the whole range of momenta. Second, the energy density stored in the initial thermal bath, $\rho_\text{DM}$, should not exceed that of the inflaton, $\rho_\text{inf}$, thus spoiling the inflationary dynamics. Since we are being agnostic about the dynamics that generates the thermal bath, we can imagine two situations depending on whether the bath formed and freezed out before of after the CMB modes exited the horizon. In the former case (thermal bath already present at CMB horizon exit) we have explicitly checked that requiring the radiation energy density to be subdominant implies temperatures so low that no effect is seen in GPP. In the latter case (the thermal bath formed and freezed out after the CMB modes exited the horizon), we require the initial condition $a_i$ to be set at the latest at the time the dominant mode $k_\star$ exits the horizon, i.e. $\rho_\text{DM}(a_i = k_\star/H_\text{inf}) < \rho_\text{inf}$:
\be\label{eq:rhoDM<rhoinf}
\rho_\text{DM}\left(a_i=\frac{k_\star}{H_\text{inf}}\right) = \frac{1}{a_i^4}\frac{\pi^2 T^4}{30} < \rho_\text{inf}=3M_P^2H_\text{inf}^2\,.
\ee
The equation above can be recast as
\al{\label{eq:thermal_bound_density}
\frac{T}{a_eH_\text{inf}} & < \frac{k_\star}{a_eH_\text{inf}} \sqrt{\frac{3\sqrt{10}}{\pi}\frac{M_P}{H_\text{inf}}}\\
& \simeq \left(\frac{k_\star/a_eH_\text{inf}}{10^{-3}}\right)\sqrt{\frac{10^{13}~\text{GeV}}{H_\text{inf}}}.
}
Of course, if the complete dynamics leading to the thermal bath is specified, the energy density of the bath can be followed precisely and the condition $\rho_\text{DM} < \rho_\text{inf}$ must be imposed in such a way that CMB physics is not spoiled. In fig.\,\ref{fig:Oh2_thermal} we highlight in yellow the band around the curves where the constraint on the initial energy density described in eq.\,\eqref{eq:thermal_bound_density} is violated. We notice that most of the low mass regime is excluded by this bound, while for very low reheating, for instance $T_\text{RH}=10^{-1}$ GeV, all points are disfavored. In gray we include bounds from super-radiance\,\cite{Cardoso:2018tly,Unal:2020jiy,Chen:2022nbb,Saha:2022hcd}. This phenomenon consists in the exponential growth of the occupation number of bosons that are gravitationally bound to astrophysical black holes, and end up depleting their spin\,\cite{Arvanitaki:2010sy,Brito:2015oca}. Other potentially relevant constraints to GPP-produced DM would be isocurvature and Lyman-$\alpha$ bounds. We argue in app.\,\ref{app:bounds}, however, that they will not exclude any portion of the parameter space shown in fig.\,\ref{fig:Oh2_thermal}.

Our results show concretely that $i)$ non Bunch--Davies initial conditions can have a huge impact on the outcome of GPP for relics and, regarding DM production, $ii)$ we can obtain the correct present abundance in a large range of previously excluded masses. For the choice of parameters shown in fig.\,\ref{fig:Oh2_thermal} we see that, depending on $T_\text{RH}$, we can obtain the correct abundance with masses that go from $m\lesssim 10^{-17}$ GeV up to $m\lesssim H_\text{inf}$, while setting Bunch--Davies initial conditions would select the value $m \simeq 10^{-13}$ GeV. This proves that precise knowledge of the initial conditions can be critical in predicting the correct abundance of spin-1 relics. 

As a final remark, we have chosen a thermal initial condition to exemplify our results, as it is a simple and physically motivated distribution. Nevertheless, our approach holds for any generic initial conditions that satisfy basic requirements such as integrability and having a sufficiently small initial energy density. If on top of that the spectrum preserves its peak on $k_\star$ as for the thermal case, analogous results should follow.

\subsection{Two-stage inflationary dynamics}\label{sec:pre_inflation}

In this section we take an alternative viewpoint, and we ask what happens when inflation is non-standard, while far in the past the vacuum state for DM was still the Bunch-Davies one. For example, if inflation consists of two epochs separated by a window of growing Hubble radius (i.e. radiation), then effectively modes that have wavelengths comparable to the horizon at the end of the first inflationary period will not correspond to Bunch-Davies solutions.

\begin{figure}[t]
\centering
\includegraphics[width=0.48\textwidth]{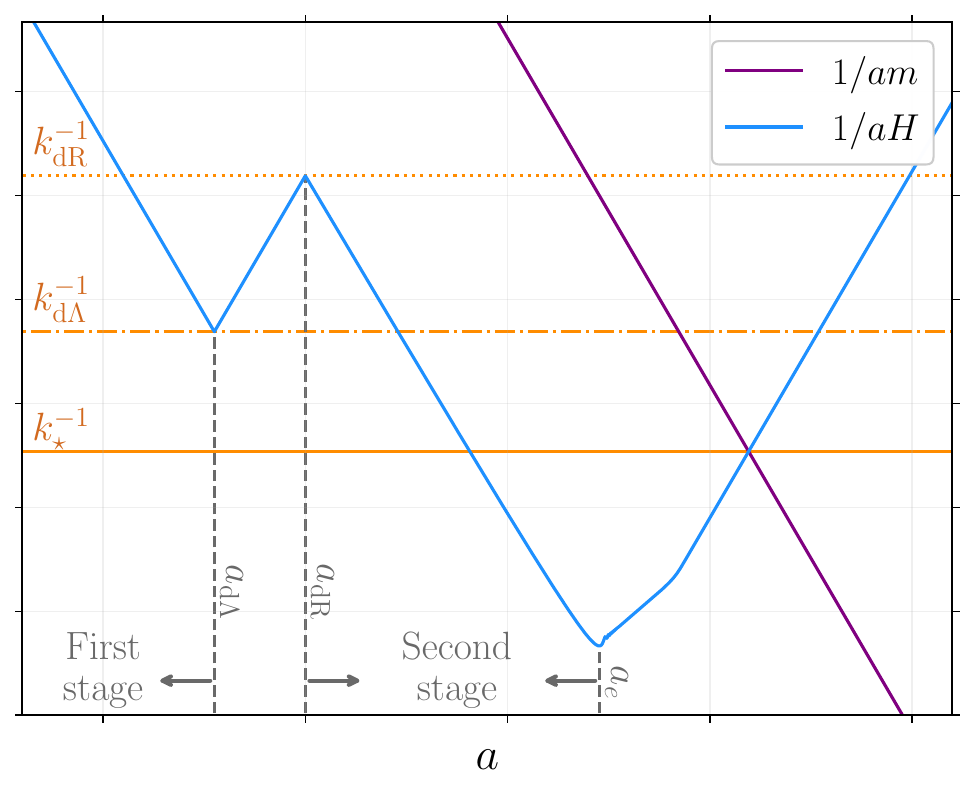}
\caption{Comoving horizon for a two-stage inflation with an intermediate radiation domination phase (solid blue). In orange we show some of the relevant comoving momentum scales and in purple we show an example of comoving Compton wavelength $1/am$. See text for details.}\label{fig:rollercoaster}
\end{figure}

Consider the situation proposed for instance in refs.\,\cite{Burgess:2002ub,Wang:2007ws}, where concrete models in which a non-standard inflationary evolution like the one shown in fig.\,\ref{fig:rollercoaster} are presented. More precisely, we consider a first period of quasi de Sitter (dS) evolution, followed by an intermediate phase -- typically either radiation or matter domination -- after which the second and final inflationary stage begins, with the usual cosmological evolution following.

This type of cosmology was described, for example, in ref.\,\cite{DAmico:2020euu}. Inflation does not have to happen ``in one go'', i.e. in a single stage of quasi dS expansion lasting the usual 60 e-folds. What is important is that the period in which CMB anisotropies are generated lasts for $10-15$ $e$-folds in order for this scenario to produce the observed spectrum. The remaining evolution can consists of dS ``bursts'' with periods of radiation/matter domination in between, provided the total duration is about 60 $e$-folds or larger. This exploits the fact that little is known about the last e-folds of inflation, which correspond to extremely short scales.

In our case, we consider the first and second period of quasi dS evolution (with Hubble parameters $H_I$ and $H_e$, respectively), to be connected by a radiation domination phase lasting $N_\text{dR}$ $e$-folds. Qualitatively similar results would follow for other types of intermediate evolution. The transition between the first stage of quasi dS and radiation phases happens at $a_{\dd\Lambda}$, while the transition between the radiation and the second stage of inflation happens at $a_\text{dR}$. The values of the scale factors $a_{\dd\Lambda}$ and $a_\text{dR}$ each define a characteristic comoving momentum scale, $k_{\dd \Lambda}$ and $k_\text{dR}$, respectively (see fig.\,\ref{fig:rollercoaster}). We suppose $k_\text{CMB} < k_\text{dR}$, with $k_\text{CMB}$ the CMB pivot scale, i.e. the generation of CMB anisotropies happens during the first stage of quasi dS evolution. With our assumptions, the Hubble parameters of the two inflationary stages are related by
\be\label{eq:NdR_definition}
H_e = H_I \left(\frac{a_{\dd \Lambda}}{a_\text{dR}} \right)^2 = H_I\, e^{-2 N_\text{dR}}\,.
\ee
We can now estimate analytically the final energy density, following the same steps that lead in the standard case to eq.\,\eqref{eq:na3_vector_standard}. We have three different cases: 
\begin{itemize}
    \item $k_\star < k_\text{dR}$ (light masses), for which
\be\label{eq:na3_vector_light}
na^3 \simeq \frac{3(a_e \sqrt{H_e\,H_I})^3}{8 \pi^2} \sqrt{\frac{H_I}{m}}.
\ee
The particle spectrum has a unique peak at $k_\star$;
    \item $k_\text{dR} < k_\star < k_{\dd \Lambda}$ (intermediate masses), for which
\be\label{eq:na3_vector_intermediate}
na^3 \simeq \frac{(a_e \sqrt{H_e\,H_I})^3}{4 \pi^2} \left(\frac{a_\text{dR}}{a_e}\right)^2 \frac{H_e}{H_I} \left(\frac{H_I}{m}\right)^{3/2}.
\ee
The particle spectrum has a unique peak at $k_\text{dR}$;
    \item $k_{\dd\Lambda} < k_\star$ (large masses), for which
\be\label{eq:na3_vector_heavy}
na^3 \simeq \frac{3(a_e H_e)^3}{8 \pi^2} \sqrt{\frac{H_e}{m}}\, .
\ee
The particle spectrum has two peaks, one at $k_\text{dR}$ and the other at $k_\star$. The quoted formula applies when the peak at $k_\star$ dominates the integral of the number density. When, on the contrary, the peak at $k_\text{dR}$ dominates, the expression for the comoving number density is well approximated by eq.\,\eqref{eq:na3_vector_intermediate}.
\end{itemize}
More details about the particle spectrum are given in app.\,\ref{app:spectrum_two_stage_inflation}. 

The abundance can now be computed as in eq.\,\eqref{eq:Omega_vector_standard} identifying $H_\text{inf}$ with the Hubble at the end of inflation, $H_e$:
\be\label{eq:Omega_vector_twostage}
\frac{\Omega_\text{vector}}{\Omega_\text{CDM}} \simeq \left(\frac{T_\text{RH}}{10^9\,\text{GeV}}\right)\left(\frac{H_e}{10^{12}\,\text{GeV}}\right)^2 \frac{m}{H_e} \frac{na^3/(a_e H_e)^3}{10^{-5}}.
\ee
The first and last case resemble quite closely eq.\,\eqref{eq:na3_vector_standard}. This is due to the fact that $k_\star$ (at which we still have at least one of the peaks in the spectrum) exits the horizon either during the first or second dS periods and then re-enters only after inflation has ended. This is what happens also in the usual computation with a single inflationary stage. On the contrary, when $k_\text{dR} < k_\star < k_{\dd \Lambda}$, the peak is at $k_\text{dR}$, causing a deformation of the spectrum and a quite different expression for the final energy density. We will show the numerical results for the spectrum later. If we want to compare eqs.\,\eqref{eq:na3_vector_light}-\eqref{eq:na3_vector_heavy} with eq.\,\eqref{eq:na3_vector_standard}, we need to decide how the Hubble parameter $H_\text{inf}$ of single-stage inflation used in the standard computation compares with $H_e$ and $H_I$. If $H_\text{inf} = H_e$, then the two-stage inflationary scenario implies an \textit{increase} in the number of gravitationally produced vectors. This can be understood from fig.\,\ref{fig:rollercoaster}, since in this case the most important modes spend more time outside the horizon. On the contrary, identifying $H_\text{inf} = H_I$, then the dominant modes spend less time outside the horizon and gravitational production is \textit{suppressed}. 

For our numerical simulations, we model both inflationary epochs using a quadratic inflationary potential, taking $\log(a_e/a_\text{dR}) \lesssim 50$ to ensure that CMB modes satisfy $k_\text{CMB} < k_\text{dR}$. Since we are now using a concrete potential for the inflaton, the approximation of constant Hubble during the inflationary stages is not necessarily very good. In fact, for the values of $\log(a_e/a_\text{dR})$ mentioned above, the Hubble parameter at the beginning of the second stage of inflation is roughly a factor $e^2$ larger then its value at the end. A similar growth would take place also in the first inflationary stage. In practice, we proceed as follows: we take $H_I$ to be the Hubble near the end of the first inflationary stage (which, given the values of $\log(a_e/a_\text{dR})$ and $N_\text{dR}$ we consider, also roughly corresponds to the Hubble at CMB scales) and $H_e$ the Hubble at the end of inflation. We then relate $H_I$ and $H_e$ correcting eq. \eqref{eq:NdR_definition} to $H_e = H_I \exp(- 2 N_\text{dR} - 2)$ to take into account the Hubble evolution in the second inflationary stage of the quadratic inflationary model.

\begin{figure*}[t]
\centering
\includegraphics[width=\textwidth]{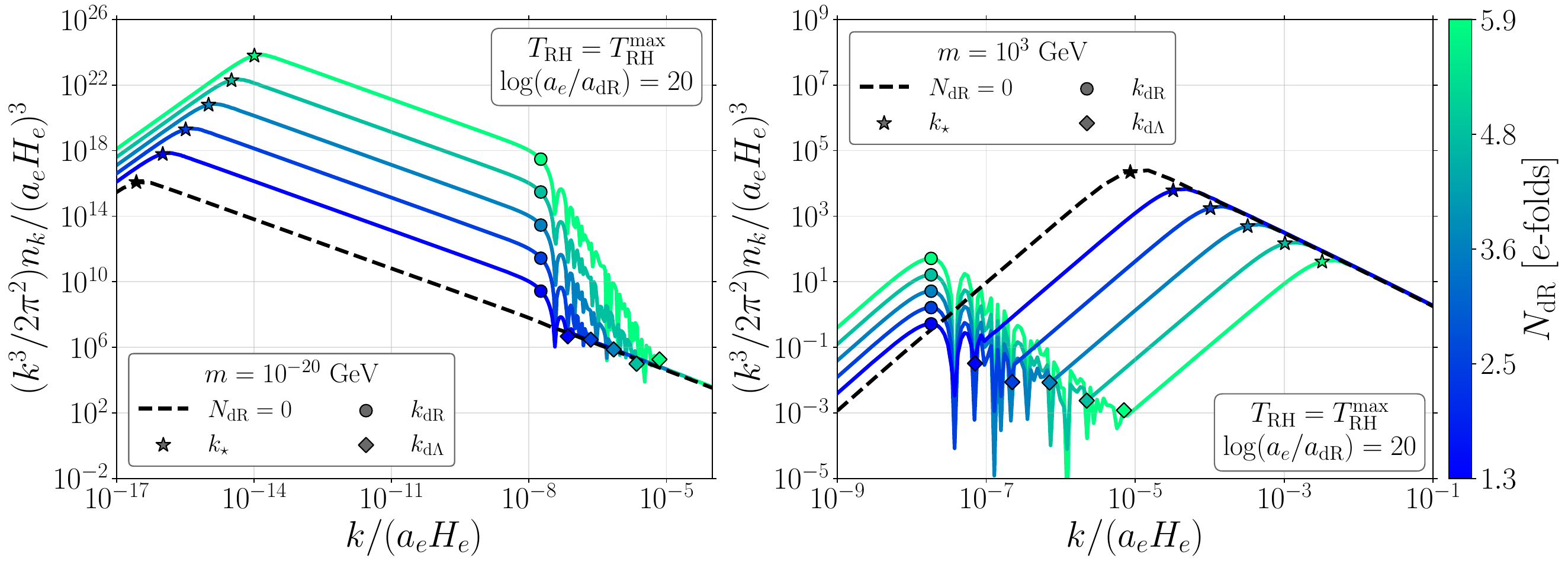}
\caption{Comoving momentum spectra for a spin-1 relic with a low mass $m=10^{-20}$ GeV ($m<m_\text{crit}$, left panel) and a high mass $m=10^{3}$ GeV ($m>m_\text{crit}$, right panel). We have fixed for both plots $T_\text{RH}=T_\text{RH}^\text{max}$, i.e. instantaneous reheating. The dashed black lines denotes results obtained with $H_e=H_I$ and $N_\text{dR}=0$. This corresponds to the usual case provided we identify $H_\text{inf} = H_e$. Colors assume an intermediate radiation phase of duration $N_\text{dR}$. The values of $k_\text{dR}$,  $k_{\dd \Lambda}$ and $k_\star$ are denoted by the circles, diamonds and stars, respectively.}\label{fig:Spectrum_preinflationary}
\end{figure*}

\begin{figure*}[t]
\centering
\includegraphics[width=0.9\textwidth]{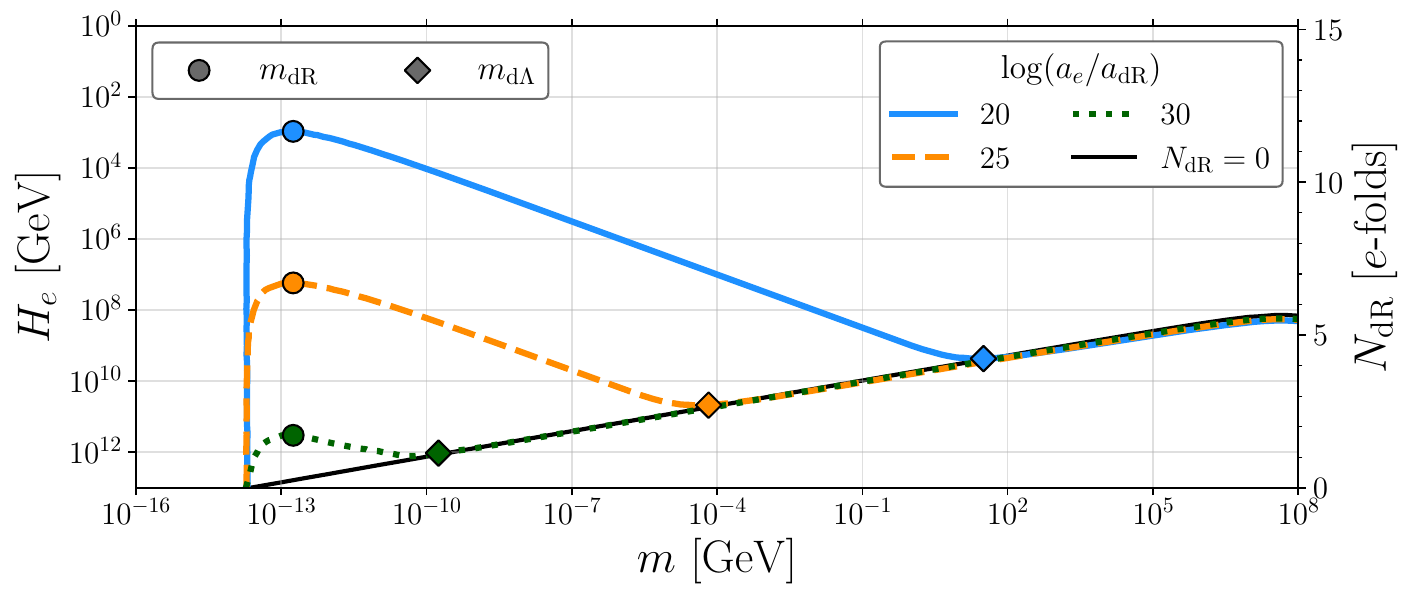}
\caption{Curves in which $\Omega_\text{vector} = \Omega_\text{CDM}$ for the two-stage inflationary case in the $m-H_e$ (or $m-N_\text{dR}$) parameter space for $\log(a_e/a_\text{dR})=20$ (solid blue), $25$ (dashed orange) and $30$ (dotted green). The black solid line represents the usual result obtained without the two-stage inflation scenario identifying $H_\text{inf} = H_e$. We also indicate with circles and diamonds the values of $m_\text{dR}$ and $m_{\text{d}\Lambda}$, see eq.\,\eqref{eq:mdR_mdLambda}.}
\label{fig:Oh2_preinflationary}
\end{figure*}

In fig.\,\ref{fig:Spectrum_preinflationary} we present the results for the normalized spectrum $(k^3/2\pi^2)n_k/(a_eH_e)^3$ as a function of the normalized comoving momentum $k/(a_eH_e)$, for $m=10^{-20}$ GeV (left panel) and $m=10^{3}$ GeV (right panel), with fixed $\log(a_e/a_\text{dR})=20$ and varying $N_\text{dR}$ according to the color code. The dashed black curve corresponds to the scenario with $H_e=H_I$ (or equivalently $N_\text{dR}=0$) corresponding to the usual single-stage inflationary dynamics with $H_\text{inf} = H_e$, and for each $H_e$ the reheating temperature is set to its maximum value $\pi^2g_*(T_\text{RH}^\text{max}){T_\text{RH}^\text{max}}^4/90M_P^2=H_e^2$. We show through the colored stars, circles and diamonds the respective values of $k_\star$, $k_\text{dR}$ and $k_{\text{d}\Lambda}$, respectively.  Furthermore, notice that both spectrum and comoving momentum are normalized with respect to the Hubble parameter of the second stage, which is the appropriate normalization required to compute the final abundance\,\eqref{eq:Omega_vector_twostage}. We reemphasize that the choice of normalization is crucial to understand how to compare with the usual case $N_\text{dR}=0$. In the left panel of fig.\,\ref{fig:Spectrum_preinflationary} for instance, while for our choice of normalization the spectrum grows the higher (lower) $N_\text{dR}$ ($H_e$) is, we see from eq.\,\eqref{eq:na3_vector_light} that the spectrum would decrease as $(H_e/H_I)^{3/2}$ had we chosen to normalize it with $H_I$. Instead, in the right plot, $k_\star\gg k_\text{dR}$ is located within the second stage of inflation, thus, with respect to $H_e$, it decreases as $\sqrt{H_e}$ according to eq.\,\eqref{eq:na3_vector_heavy}; normalizing it with respect to $H_I$ would just sharpen this effect.

The two panels in fig.\,\ref{fig:Spectrum_preinflationary} show qualitatively different features depending on the value of the masses. More precisely, we choose a very low value $m=10^{-20}$ GeV (left plot) for which $k_\star\ll k_\text{dR},k_{\text{d}\Lambda}$, and a very large value $m=10^3$ GeV that satisfies the opposite relation $k_\star\gg k_\text{dR},k_{\text{d}\Lambda}$. In the left plot (low mass), as anticipated, we observe that the peak of the spectrum remains at $k_\star$. For comoving momenta $k>k_{\dd\Lambda}$ the spectrum converges to the case $N_\text{dR}\to 0$, while for $k<k_\text{dR}$ it has the exact same shape, but shifted above as the modes evolve for a longer time. These feature can be read directly from the analytical estimates presented in app.\,\ref{app:spectrum_two_stage_inflation}, in particular the peak value of the spectrum changes as $(k^3/2\pi^2)n_k|_{k=k_\star}/(a_eH_e)^3\simeq (H_I/H_e)^{3/2}\sqrt{H_I/m}$, thus for lower $H_e$ (higher $N_\text{dR}$), the higher the peak is. The most interesting feature comes for modes with $k_\text{dR}<k<k_{\dd\Lambda}$ , i.e. the modes that exit and enter the horizon before the last stage of inflation, which arrive at the beginning of the second stage of inflation with non Bunch--Davies initial conditions. Here the spectrum behaves as $k^{-5}$, which is steeper than the usual case (see app.\,\ref{app:spectrum_two_stage_inflation}). Moreover, this part of the spectrum is oscillating, a feature that appears because the modes exit, enter and re-exit the horizon. 

For the plot on the right, choosing a much larger mass, $m=10^3$ GeV, we observe the same oscillatory features for modes in the range $k_\text{dR}<k<k_{\dd\Lambda}$ and also an increase of the spectrum for $k<k_\text{dR}$, while maintaining the same slope. However, different from the low-mass case, a second peak emerges at $k_\text{dR}$ (see app. \ref{app:spectrum_two_stage_inflation}), which can surpass the one at $k_\star$ for sufficiently large $N_\text{dR}$. We stress once more that the plots, and our discussion, depends strongly on the interpretation $H_\text{inf} = H_e$. Had we chosen to normalize the spectra to $(a_e H_I)^3$, identifying the dashed black line with $H_\text{inf} = H_I$, then the spectrum would always be decreased with respect to the standard case.

As is clear from the discussion, the qualitative behavior of the spectrum (i.e. the number and location of the peaks) is completely determined by the vector mass since, for fixed inflationary parameters, this gives the position of $k_\star$ with respect to $k_\text{dR}$ and $k_{\dd\Lambda}$. It is thus convenient to introduce
\al{\label{eq:mdR_mdLambda}
m_\text{dR} & \equiv H_e \left(\frac{a_\text{dR}}{a_e}\right)^2,\\
m_{\dd\Lambda} & \equiv \frac{3}{2}\frac{H_I^2}{H_e} \left(\frac{a_\text{dR}}{a_e}\right)^2 = \frac{3}{2}\left(\frac{H_I}{H_e}\right)^2 m_\text{dR},
}
the masses for which $k_\star = k_\text{dR}$ and for which the peaks at $k_\text{dR}$ and $k_\star$ give the same contribution to $na^3$, respectively. For simplicity, the expressions are given for instantaneous reheating (i.e. setting $H_\text{inf} = H_\text{RH}$ in eq.\,\eqref{eq:kstar}), but they can be easily generalized. We use $k_\text{dR} = a_\text{dR} H_e$ and $k_{\dd\Lambda} = a_{\dd\Lambda} H_I$. As we saw before, when $m < m_\text{dR}$ we have a unique peak at $k_\star$ while, for $m > m_\text{dR}$, we always have at least a peak at $k_\text{dR}$. This means that we can have power injected in the spectrum at relatively small momenta. If this happens at CMB scales, strong limits from isocurvature perturbations must be taken into account. Such bounds are avoided requiring $k_\text{CMB} < k_\text{peak} \equiv \text{min}(k_\text{dR}, k_\star)$. A dedicated discussion can be found in app.\,\ref{app:isocurvature}, in which we show that we have no isocurvature limits for the parameters considered in the numerical analysis.

Finally, we present in fig.\,\ref{fig:Oh2_preinflationary} the curves for which $\Omega_\text{vector} = \Omega_\text{CDM}$ as a function of the mass of the vector and the duration of the intermediate radiation phase or, equivalently, the Hubble at the end of the second stage of inflation. In the figure, we show the results for $\log(a_e/a_\text{dR})=20$ (solid blue), $25$ (dashed orange) and $30$ (dotted green), and in solid black the $N_\text{dR} = 0$ case. This corresponds to the usual case of a single inflationary stage provided we identify $H_\text{inf} = H_e$. For the two-stage inflationary case, we take $H_I=10^{14}$ GeV. For all choices of $H_e$, we consider instantaneous reheating for simplicity, but we do not expect qualitative changes for finite reheating.  In all cases with a two-stage inflation, the curves grow vertically up to a certain value of $H_e$, after which they start slowly decreasing until at some point they converge to the usual prediction. This behavior can be easily understood inserting eqs.\,\eqref{eq:na3_vector_light}-\eqref{eq:na3_vector_heavy} into eq.\,\eqref{eq:Omega_vector_twostage}. For small masses, $\Omega_\text{vector}$ is independent from $H_e$, which is reflected in the vertical part of the curves in fig.\,\ref{fig:Oh2_preinflationary}. For intermediate masses, imposing $\Omega_\text{vector} = \Omega_\text{CDM}$ requires $H_e^{-1} \propto m^{-1/2}$ while, for larger masses, it requires $H_e^{-1} \propto m^{1/4}$. This behavior corresponds to the one observed in the figure, in which we also show the values of $m_\text{dR}$ (colored circles) and $m_{\dd\Lambda}$ (colored diamonds). We see that our analytical estimates reproduce well the masses for which a change of behavior occurs.

We conclude this section with two final comments. First, as already mentioned, no region in fig.\,\ref{fig:Oh2_preinflationary} is excluded by isocurvature perturbations limits for our choice of parameters. The same is true also for other possible limits that could be imposed on the parameter space (e.g. Lyman-$\alpha$ and white noise constraints). A detailed discussion can be found in app.\,\ref{app:bounds}. Finally, we observe that the overall effect of the two-stage inflationary scenario is to allow, given a certain vector mass, to obtain the correct abundance for different values of the early universe parameters $H_e$ and/or $T_\text{RH}$, even much smaller than what happens in the standard case. However, as can be seen from fig.\,\ref{fig:Oh2_preinflationary},  this happens only for masses $m \gtrsim 6\times 10^{-6}$ eV, which corresponds to the mass that ensures that the correct abundance is obtained for $H_e \simeq 10^{14}$ GeV\,\cite{Graham:2015rva,Ahmed:2020fhc,Kolb:2020fwh,Redi:2022zkt}. Smaller masses would reproduce the correct abundance only for larger values of $H_e$, experimentally excluded. As in the standard computation with a single inflationary stage, smaller masses can be relevant only admitting that the vector does not constitute the whole of DM.


\section{Conclusions}\label{sec:conclusions}

The evolution of quantum fields during cosmological evolution allows for the production of relics, in particular DM, in an elegant and physically interesting way, without having to resort to any additional interactions of the relics other than  gravity. A precise calculation of the resulting abundance, 
crucial for the correct determination of the DM energy density, is nonetheless based on a number of assumptions. In this paper, we have relaxed one of these assumptions: instead of taking no particles at the beginning of inflation (Bunch--Davies vacuum), we allow for an initial population of relics. Our results thus generalize the description of GPP with arbitrary initial conditions.

We have shown that GPP in the presence of non Bunch--Davies initial conditions does not simply amount to adding the initial abundance to the usual outcome of GPP with Bunch--Davies initial conditions. Instead, as illustrated by eqs.\,\eqref{eq:nk_1}-\eqref{eq:nk_2}, the quantum mechanical nature of the relics predicts that the final number density is either "stimulated" or ``blocked'' due to terms that are proportional to both initial density and Bogoliubov coefficients. While it was known in the literature that non-trivial initial conditions could give rise to interesting effects on the GPP\,\cite{Parker:1968mv,Parker:1969au,Parker:1971pt}, to the best of our knowledge it was neither explored in full generality nor studied in the context of relics. For initial conditions that affect relatively small $k$, a general conclusion that can be drawn from eqs.\,\eqref{eq:nk_1}-\eqref{eq:nk_2} is that GPP is almost unaffected by initial conditions for fields whose only break of conformality is due to their mass (conformally coupled scalar, fermion, transverse polarizations of spin-1). This is easily explained: for these fields, there is basically no GPP until the modes cross the comoving Compton wavelenght and the mass becomes important. This implies that there is almost no superposition between the $k$'s affected by initial conditions and the peak of GPP. The final abundance computed in the usual way is thus robust for these fields. For the remaining cases, it is known that a minimally coupled scalar is excluded by isocurvature perturbation limits in the Bunch--Davies case; this conclusion is not changed by different initial conditions. We are thus left with the longitudinal mode of spin-1 fields, for which GPP starts well before the mode crosses the Compton wavelength and that requires a quantitative analysis.

To do so, we have chosen two different scenarios, the first with a thermal initial distribution, remaining agnostic about its origin, and the second where we follow explicitly a two-stage inflationary dynamics in which we have two periods of almost de Sitter expansion with an intermediate stage of radiation domination.
Both in the thermal (figs.\,\ref{fig:Spectrum_thermal}, \ref{fig:Oh2_thermal}) and in the two-stage inflation
(figs.\,\ref{fig:Spectrum_preinflationary}, \ref{fig:Oh2_preinflationary}) scenarios, the spectrum shows important differences with respect to the usual one, thus impacting also the abundance significantly. In the thermal case, we see from fig.\,\ref{fig:Oh2_thermal} that, for fixed $H_\text{inf}$ and $T_\text{RH}$ and different values of $T$, the correct abundance can be obtained for masses that can both be smaller and larger than the value obtained with Bunch--Davies initial conditions. In the two-stage inflationary case, on the other hand, we see from fig. \ref{fig:Oh2_preinflationary} that only masses $ m \gtrsim 6\times 10^{-6}$ eV are allowed, where the quoted number is the one obtained for an inflationary Hubble parameter at CMB scales as high as $10^{14}$ GeV. Still, for these masses the correct abundance can be obtained for values of $H_e$ and $T_\text{RH}$ that can be much smaller than those appearing in the standard computation.

Our work can be extended in several directions. First, we have examined two examples of non-trivial initial conditions, but one could systematically list different possibilities to find out if GPP is more sensitive to a given class of initial conditions than others. Moreover, while our analysis was restricted up to spin-1 particles, one could also evaluate how higher-spin or exotic particles are affected by non Bunch--Davies initial conditions. Finally, an aspect to be explored is the inclusion of more interactions, either among the relics or with SM particles, that in particular could be the ones responsible for generating non-trivial initial conditions. We leave these points for future work.


\subsection*{Acknowledgments}
We thank the participants of MITP scientific program ``Theory Facing Experiment on the Dark Matter and Flavor Puzzles'' for asking questions about cosmological gravitational particle production that lead to this project, in particular Tony Gherghetta and Lisa Randall. 
The work of EB is partly
supported by the Italian INFN program on Theoretical Astroparticle Physics (TAsP). GMS acknowledges financial support from ``Fundação de Amparo à Pesquisa do Estado de São Paulo" (FAPESP) under contracts 2020/14713-2 and 2022/07360-1, from the Deutsche Forschungsgemeinschaft under Germany’s Excellence Strategy – EXC 2121
“Quantum Universe”, as well as from the grant 491245950. This project received funding from the European Union’s Horizon Europe research and innovation
program under the Marie Skłodowska-Curie Staff Exchange grant agreement No 101086085
– ASYMMETRY. The work of AT is supported by the Italian Ministry of University and Research (MUR) through the PRIN
2022 project n. 20228WHTYC (CUP:I53C24002320006).
\medskip

\appendix
\section{Conventions and useful formulas}\label{app:conventions_and_fields}
We collect in this appendix useful more details on the derivation of the equations used in sec.\,\ref{sec:review}.
\subsection{Spin-0 field}
\begin{table}[]
    \centering
    \begin{tabular}{c||c|c}
        $\phi_{k}$    &  de Sitter  & Radiation  \\ \hline \hline 
     $k \gg a H, am$ &   $c_\pm e^{\pm i k \eta} $ &  $c_\pm e^{\pm i k \eta} $ \\
     $aH \gg k \gg am$   & $c_\pm a^{-\frac{1}{2} \left(1 \pm \sqrt{1- 8 \Delta\xi}\right)}$ & $c_\pm a^{\frac{1}{2} \pm \frac{1}{2}} $ \\
     $aH \gg am \gg k$ &  $c_\pm a^{-\frac{1}{2} \left(1 \pm \sqrt{1- 8 \Delta\xi}\right)}$  & $c_\pm a^{\frac{1}{2} \pm \frac{1}{2}} $ \\
     $am \gg k, aH$  & / & $c_\pm \frac{e^{\pm i \hat{\Omega}_k(\eta) }}{\sqrt{a}} $
    \end{tabular}
    \caption{Approximate solutions $\phi_{k}$ of the mode equation\,\eqref{eq:EoM_harmonic_oscillator} with frequency\,\eqref{eq:frequency_scalar} in different regimes. For simplicity, we consider only a pure de Sitter or radiation dominated universe. Matching these solutions allows to compute the number of gravitationally produced particles in the case of instantaneous reheating. We have defined $\Delta\xi \equiv 6 \xi -1$ and $\hat{\Omega}_k = \int^\eta d\eta' \omega_k(\eta')$. We do not present the solution for $am \gg k, aH$ in the de Sitter regime because, under our assumptions, this case is never realized. Observe that, during radiation domination, $a''=0$, so that the solutions are independent from $\xi$.}
    \label{tab:solution_scalar}
\end{table}
We start from the action 
\be
S = \int d^4x \sqrt{-g} \frac{1}{2} \bigg[g^{\mu\nu} \partial_\mu \varphi \partial_\nu \varphi - m^2 \varphi^2 + \xi\,R\,\varphi^2 \bigg],
\ee
where $R$ is the Ricci scalar. If we now use conformal time and apply the field redefinition $\phi =a \varphi$ to make the kinetic term of $\phi$ canonically normalized (Weyl rescaling), we end up with the action
\be
S = \int d\eta \,d^3x\,\frac{1}{2} \bigg[(\phi')^2 + (\nabla \phi)^2 - m_\text{eff}^2 \phi^2  \bigg],
\ee
where 
\be\label{eq:frequency_scalar}
m_\text{eff}^2  =  a^2 m^2 - a^2 \left(\xi - \frac{1}{6} \right) R .
\ee
From this action we obtain an EoM in the form of eq.\,\eqref{eq:EoM_harmonic_oscillator},
\be
\phi_k'' + \omega_k^2 \, \phi_k = 0, ~~~~ \omega_k^2 = k^2 + m_\text{eff}^2,
\ee
with $k$ the comoving momentum and $\phi_k$ the Fourier transform of $\phi(x)$, and an Hamiltonian 
\be
\mathcal{H} = \int d^3x \frac{1}{2} \bigg[(\phi')^2 + (\nabla \phi)^2 + m_\text{eff}^2 \phi^2  \bigg].
\ee
The field $\phi$ can be quantized as in eq.\,\eqref{eq:quantum_field}. If we do so in the expression for $\mathcal{H}$ just computed, we obtain eq.\,\eqref{eq:H}. 
Requiring canonical equal-time commutation relations between $\phi(x)$ and its conjugate momentum $\pi(x) = \phi'(x)$ implies that the mode functions must satisfy the normalization condition 
\be
u_k u_{k}'^\star - u_{k}^\star u_k' = i.
\ee
We can now ask ourselves what happens after the Bogoliubov transformation of eq.\,\eqref{eq:Bogoliubov_mode_func} is applied. As explained in the main text, the commutation relations $[a_k, a_q^\dagger] = (2\pi)^3 \delta^{(3)}(\bm{k} - \bm{q})$ are preserved only if the Bogoliubov coefficients satisfy $|\alpha_k|^2 - |\beta_k|^2 =1 $. 
The Bogoliubov coefficients can be computed explicitly in terms of $u_k$ and $w_k$ using the orthogonality of mode functions under the (conserved) scalar product of the Klein-Gordon equation \cite{Birrell:1982ix}:
\be\label{eq:alpha_beta_scalar}
\alpha_k = - i (w_k\, u_{-k}'^\star - w_k'\, u_{-k}^\star), ~~~~~ \beta_k = -i  ( w_k\,u_k' - u_k\, w_k').
\ee
This equation allows to compute $\beta_k$, and hence the total number of particles produced, once we fix the mode functions $u_k$ and $w_k$. As in the main text, the most useful choice for us is to compute $u_k$ as the solution of eq.\,\eqref{eq:EoM_harmonic_oscillator} with appropriate initial conditions, while the functions $w_k$ can be chosen to select the adiabatic vacuum as in eq.\,\eqref{eq:WKB}. Notice that computing $|\beta_k|^2$ with eq.\,\eqref{eq:alpha_beta_scalar} gives the same result obtained comparing eqs.\,\eqref{eq:N_1} and\,\eqref{eq:N_2}, confirming the consistency of our approaches. Solutions of eq.\,\eqref{eq:EoM_harmonic_oscillator} are shown in tab.\,\ref{tab:solution_scalar}. They are computed approximating inflation as a pure de Sitter phase of expansion and taking instantaneous reheating, i.e. inflation is followed immediately by the usual radiation domination phase. Using tab.\,\ref{tab:solution_scalar} we can find the transfer function $T(k,\eta)$. For instance, for a minimally coupled scalar ($\xi = 0$) we have
\be\label{eq:transfer_minimally_coupled}
T(k, \eta) \propto \left\{
\begin{array}{ccc}
    \text{const} & ~~~ & k \gg aH,\,am   \\
    a^2 &  ~~~ & aH \gg k,\, am \\
    a^{-1} & ~~~ & am \gg k,\,aH
\end{array}\right.
\ee
In the case of Bunch--Davies initial conditions, we can combine eqs. \eqref{eq:power_spectrum_transfer_func}, \eqref{eq:Pk_initial_BunchDavies} and \eqref{eq:transfer_minimally_coupled} to obtain $(k^3/2\pi^2) n_k$. The result scales as $k^0$ for $k<k_\star$ and $k^{-1}$ for $k>k_\star$. According to app. \ref{app:isocurvature}, this case is excluded by isocurvature limits.

\subsection{Spin-1/2 field}\label{app:spin1/2}
We start from the action of a fermion field in a curved background\,\cite{Parker:1971pt,Chung:2011ck,Adshead:2015kza,Ema:2019yrd,Koutroulis:2023fgp}
\be
S = \int d\eta d^3x \, \sqrt{-g} \,\bigg[ \overline{\psi} (i\,e^\mu_a \,\gamma^a \,\nabla_\mu - m ) \psi \bigg],
\ee
where $e^\mu_a$ is the vierbein, defined by $\eta_{ab} = e_a^\mu e_b^\mu g_{\mu\nu}$ and the covariant derivative is defined by 
\be
\nabla_\mu \equiv \partial_\mu + \frac{1}{2}\omega_{\mu a b} \Sigma^{ab},
\ee
with spin connection given by
\be
\omega_\mu^{ab} \equiv e^{a\nu} \left(\partial_\mu e^b_\nu - \gamma^\lambda_{\mu\nu} e^b_\lambda\right),
\ee
and $\Sigma^{ab} \equiv [\gamma^a, \gamma^b]/4$ the generator of Lorentz transformations in spinor representation. For a FLRW metric in conformal time, we have
\be
\omega_0^{ab} = 0 = \omega_i^{jk}, ~~~~~ \omega_i^{0j} = \frac{a'}{a}\,\delta^j_i,
\ee
so that with a field redefinition $\chi = a^{3/2} \psi$ (Weyl rescaling) the fermion action becomes
\be
S = \int d^3x\,d\eta\, \bar{\chi} (i \gamma^b \partial_b - a m) \chi.
\ee
The EoM that follows from this action is simply
\be
(i \gamma^b \partial_b - a m)\chi = 0.
\ee
We can now proceed as usual, but at this point, for simplicity, we assume that the fermion is Majorana: we write
\be
\chi = \sum_{\lambda = \pm 1} \int \frac{d^3k}{(2\pi^3)} \left[u_{k\lambda}(\eta) a_{k\lambda} e^{i \bm{k} \cdot \bm{x}} + v_{k\lambda}(\eta) a_{k\lambda}^\dagger e^{-i \bm{k} \cdot \bm{x}}  \right]  ,
\ee
where, as usual, $v_{k\lambda} = - i \gamma^2 u_{k\lambda}^\star$. Using the Dirac representation of the $\gamma$ matrices, the spinor $u_{k\lambda}$ must satisfy the equation
\be\label{eq:EoM_u}
i \gamma^0 u_{k\lambda}' - \bm{\gamma} \cdot \bm{k} \, u_{k\lambda} - a m  \, u_{k\lambda}  = 0.  \\
\ee
We write $u_{k\lambda} = (u_{Lk\lambda}, u_{Rk\lambda})^T$ and decompose $u_{L,R}$ as
\be
\begin{aligned}
u_{Lk\lambda} & = x_{k\lambda}(\eta) \,h_{k\lambda} , \\
u_{Rk\lambda} & = y_{k\lambda}(\eta) \,h_{k\lambda} ,
\end{aligned}
\ee
where $x_{k\lambda}$ and $y_{k\lambda}$ are simple functions and all the spinor indices are carried by the helicity eigenvectors $h_{k\lambda}$, that satisfy
\be
(\bm{\sigma} \cdot \hat{\bm{k}}) h_{k\lambda} = \lambda h_{k\lambda} .
\ee
The helicity eigenvectors satisfy the identities
\be
h_{-k\lambda} = \xi_{k\lambda} h_{k-\lambda}, ~~~~ \sigma^2\,h_{k\lambda}^\star = i \lambda h_{k-\lambda},
\ee
where the phase $\xi_{k\lambda}$ can always be chosen to be $\xi_{k\lambda} = - \lambda e^{i \lambda \phi}$, with $\phi$ the azymuthal angle that identifies the momentum $\bm{k}$ direction. 
In terms of the functions $x_{k\lambda}$ and $y_{k\lambda}$, eq.\,\eqref{eq:EoM_u} becomes
\be\label{eq:EoM_y_x}
\begin{aligned}
x_{k\lambda}' & = -i a m\, x_{k\lambda} - i \lambda k \, y_{k\lambda},\\
y_{k\lambda}' & =  i a m \, y_{k\lambda}  -i  \lambda k \, x_{k\lambda} , 
\end{aligned}
\ee
which can be rewritten as two second order differential equations:
\be
\begin{aligned}
x_{k\lambda}'' + (\omega_k^2 + i a' m) x_{k\lambda} & = 0 , \\
y_{k\lambda}'' + (\omega_k^2 - i a' m) y_{k\lambda} & = 0 , \\
\end{aligned}
\ee
where $\omega_k^2 = k^2 + a^2 m^2$. 
These are equations of the harmonic oscillator type, but with intrinsically complex frequency square. Requiring the equal-time anticommutation relations $[ \chi(\eta, \bm{x}), \pi(\eta, \bm{y})]_+ = i \delta^{(3)}(\bm{x} - \bm{y})$ to be true, we discover that the mode functions $x_{k\lambda}$ and $y_{k\lambda}$ must satisfy
\be
|x_{k\lambda}|^2+  |y_{k\lambda}|^2 = 1 ~~~~~~~ \text{(no sum over $\lambda$)}.
\ee
We can use eq.\,\eqref{eq:EoM_y_x} to eliminate $y_{k\lambda}$ in favor of $x_{k\lambda}$ and $x_{k\lambda}'$, obtaining
\be\label{eq:normalization_fermion}
|x_{k\lambda}'|^2 + \omega_k^2 |x_{k\lambda}|^2 + i a m (x_{k\lambda} x_{k\lambda}^{\star \prime} - x_{k\lambda}^\star x_{k\lambda}' ) = k^2 .
\ee
We can now compute the Hamiltonian. In terms of the field $\chi$, the on-shell Hamiltonian is simply given by
\be
\mathcal{H} = \int d^3x \, i\,\chi^\dagger \chi'.
\ee
This is the same expression as in Minkowski space, since using conformal time we have the same action. Once we use the field decompositions used so far, it is a simple matter of algebra to show that
\be
\begin{aligned}
    \mathcal{H} & = \sum_{\lambda = \pm 1} \int \frac{d^3k}{(2\pi)^3} \frac{1}{2}\bigg\{\Omega_{k\lambda} \left[ a_{k\lambda}^\dagger a_{k\lambda} - a_{k\lambda} a_{k\lambda}^\dagger \right] \\
    & \quad \quad + \mathcal{F}_{k\lambda} a_{k\lambda} a_{-k\lambda} + h.c. \bigg\},
\end{aligned}
\ee
with functions defined via
\be\label{eq:Omega_k}
\begin{aligned}
\Omega_{k\lambda} &  \equiv  \frac{|x_{k\lambda}'|^2 + \omega_k^2 |x_{k\lambda}|^2 - \omega_k^2}{a m} , \\
\mathcal{F}_{k\lambda} & \equiv    \frac{\xi_{k\lambda}}{\lambda k} \bigg[  (x_{k\lambda}')^2 + \omega_k^2 (x_{k\lambda})^2 \bigg],
\end{aligned}
\ee
where we have used eqs.\,\eqref{eq:EoM_y_x} and\,\eqref{eq:normalization_fermion} to simplify the expression. This is precisely the form quoted in eq.\,\eqref{eq:H}, once we suppress the helicity indices. As a useful cross-check that our formulas are correct, let us go back to the Minkowski limit $a \to 1$. In this case, $x_{k\lambda}$ must be of the form $\sqrt{\omega_k +m}/\sqrt{2\omega_k} e^{- i\omega_k \eta}$ to match the usual flat space normalization for the 4-spinor $u_{k\lambda}$. With this choice, it is simple to show that $\Omega_k = \omega_k$ and $\mathcal{F}_k = 0$, as expected. 

We now turn to the effects of a Bogoliubov transformation. First of all, we stress again the importance of the phase $\xi_{k\lambda}$ in eq.\,\eqref{eq:Bogoliubov}, that guarantees the absence of singularities in the transformation. At the level of spinors we have
\al{
\hat{u}_{k\lambda} & = \alpha_k u_{k\lambda} + \beta_k \, \xi_{-k\lambda}\, v_{-k\lambda}, \\
\hat{v}_{-k\lambda} & = \alpha_k^\star v_{-k\lambda} + \beta_k^\star \xi_{k\lambda}^\star u_{k\lambda},
}
where the hat denotes the spinors obtained after the Bogoliubov transformation. Equivalently, at the level of $x_{k\lambda}$ and $y_{k\lambda}$ functions we have
\al{\label{eq:Bogoliubov_x_y}
\hat{x}_{k\lambda} & = \alpha_k x_{k\lambda} +  \lambda \beta_k y_{k\lambda}^\star, \\
\hat{y}_{k\lambda} & = \alpha_k y_{k\lambda} -  \lambda \beta_k x_{k\lambda}^\star. \\
}
An explicit form for the Bogoliubov coefficients can be obtained using the (conserved) scalar product for the Dirac equation. We obtain
\al{\label{eq:alpha_beta_fermion}
\alpha_k & = \frac{i (y_{k\lambda}^\star  \hat{x}_{k\lambda}' - \hat{x}_{k\lambda}  y_{k\lambda}^{\star \prime} )}{{k\lambda}}, \\
\beta_k & =\frac{i (\hat{x}_{k\lambda} x_{k\lambda}' - x_{k\lambda} \hat{x}_{k\lambda}')}{k} .
}
We observe that the $|\beta_k|^2$ obtained squaring eq.\,\eqref{eq:alpha_beta_fermion} coincides with the one obtained comparing eqs.\,\eqref{eq:N_1} and\,\eqref{eq:N_2}. 

Finally, the expectation value of the Hamiltonian can be written as
\be
\langle \mathcal{H} \rangle = \frac{1}{2}\int \frac{dk}{k} \mathcal{P}_{\chi}
\ee
if we define the power spectrum as 
\be\label{eq:power_spectrum_fermions}
\langle i\, \chi_k^\dagger(\eta) \chi_q'(\eta) \rangle = (2\pi)^3 \delta^{(3)}(\bm{k} - \bm{q}) \frac{2\pi^2}{k^3} \mathcal{P}_\chi(k, \eta).
\ee
Unlike in the bosonic case, one of the two fields appears here as a derivative. Using the explicit spinor function decomposition $\chi_k = u_k a_k + v_{-k} a_{-k}^\dagger$ and the EoM, we obtain
\be
\frac{2\pi^2}{k^3}\mathcal{P}_\chi(k, \eta) = \Omega_k \left( 2 \langle N_k^{in} \rangle - V \right) + \mathcal{F}_k\, \langle C_k^{in} \rangle + c.c.,
\ee
with $\Omega_k$ and $\mathcal{F}_k$ defined in eq.\,\eqref{eq:Omega_F_fermions}. The solutions to the equations of motions, as the transfer function, are well approximated by those of a conformally coupled scalar.

\subsection{Spin-1 field}\label{app:spin-1}
\begin{table}[]
    \centering
    \begin{tabular}{c||c|c}
        $u_{k}$    & de Sitter &  Radiation  \\ \hline \hline 
     $k \gg a H, am$ &  $c_\pm e^{\pm i k \eta}$ & $c_\pm e^{\pm i k \eta}$ \\
     $aH \gg k \gg am$ & $c_1\,a  +\frac{c_2}{a^2} $ &  $ c_1 + c_2\,a$ \\
     $aH \gg am \gg k$ & $c_1 + \frac{c_2}{a}$ &  $c_1 a + c_2$ \\
     $am \gg k, aH$ & / & $\frac{c_1 e^{-i m t} + c_2 e^{imt}}{\sqrt{a}}$
    \end{tabular}
    \caption{As in table\,\ref{tab:solution_scalar}, but for the solution  $u_{k}$ of the mode equation\,\eqref{eq:EoM_harmonic_oscillator} with frequency\,\eqref{eq:frequency_vector}. }
    \label{tab:solution_vector}
\end{table}
We start from the action\,\cite{Graham:2015rva,Ema:2019yrd,Kolb:2020fwh,Ahmed:2020fhc,Arvanitaki:2021qlj,Redi:2022zkt,Capanelli:2024pzd,Capanelli:2024rlk}
\be
S = \int d^3x d\eta\, a^4 \bigg[-\frac{1}{4} g^{\mu\alpha} g^{\nu\beta} F_{\mu\nu} F_{\alpha\beta} - \frac{m^2}{2} g^{\mu\nu} A_\mu A_\nu \bigg],
\ee
where we take the abelian case $F_{\mu\nu} = \partial_\mu A_\nu - \partial_\nu A_\mu$ and we ignore possible non-minimal couplings to the curvature (see\,\cite{Kolb:2020fwh,Capanelli:2024pzd,Lebedev:2025snd} for a discussion of their effects). The discussion of the action is easier in momentum space, to which we switch from now on. The time component $A_k^0$ is non-dynamical and can be integrated out. Decomposing the remaining degrees of freedom into transverse and longitudinal components as $\bm{A}_k = \bm{A}^T_k + \hat{\bm{k}} A_k^L$, with $\bm{k} \cdot \bm{A}_k^T = 0$, we obtain the two independent actions
\be
\begin{aligned}
    S_T & = \int \frac{ d\eta\,d^3k}{(2\pi)^3} \frac{1}{2}\bigg[|\bm{A}_{Tk}'|^2 - \left(k^2 +a^2 m^2\right) |\bm{A}_{Tk}|^2 \bigg] , \\
    S_L & = \int\frac{d\eta\,d^3k}{(2\pi)^3} \frac{1}{2} \bigg[\frac{a^2 m^2}{k^2 + a^2 m^2} |A_{Lk}'|^2 - a^2 m^2 |A_{Lk}|^2  \bigg] .
\end{aligned}
\ee
We see that the action for the transverse modes is the same as the one of a conformally coupled scalar field ($\xi=1/6$), so that from now on we will focus only on the longitudinal degree of freedom. The kinetic term of $A_{Lk}$ can be made canonic via the field redefinition $A_{Lk} = \frac{\sqrt{k^2 + a^2 m^2}}{am} u_k$ and the EoM for $u_k$ read
\be
u_k'' + \omega_k^2\,u_k = 0,
\ee
with squared frequency
\al{\label{eq:frequency_vector}
\omega_k^2 & = k^2 + a^2 m^2 - \frac{k^2}{k^2+a^2 m^2} \left(\frac{a''}{a} - \frac{3 a^4 m^2 H^2}{k^2 +a^2 m^2}\right).
}
This EoM is again of the form of eq.\,\eqref{eq:EoM_harmonic_oscillator}. Quantization now can proceed as in the case of the scalar field. In particular, all considerations about the conditions imposed by the consistency of the Bogoliubov transformation apply exactly in the same way to the vector case. 

Exact analytical solutions of the EoM are not available, but approximate expressions can be found. These have been discussed in detail in\,\cite{Graham:2015rva,Kolb:2020fwh,Ahmed:2020fhc} and are reported in tab.\,\ref{tab:solution_vector}. Using these solutions, we obtain the following transfer function:
\be\label{eq:transfer_function_vector}
T(k, \eta) \propto \left\{
\begin{array}{ccc}
    \text{const} & ~~~ & k \gg aH,\,am   \\
    a^2 &  ~~~ & aH \gg k \gg am \\
    \text{const} & ~~~ & aH \gg am \gg k \\
    a^{-1} & ~~~ & am \gg k,\,aH
\end{array}\right.
\ee
where we have maintained only the overall $a$ scaling, averaging over oscillations.

\section{Coherent states}\label{app:coherent}

In this appendix we discuss coherent states as possible non Bunch--Davies initial conditions. In particular, we will show that they cannot produce any effect in GPP, thus resulting in the same abundance as in the Bunch--Davies scenario.

We begin with the definition of a coherent state $\ket{\psi}=\ket{z}$:
\be\label{eq:coherent}
a_k\ket{z} = z(\bm k)\ket{z},
\ee
where $z(\bm k)$ is a function of $\bm k$. The relevant quantities in eqs.\,\eqref{eq:N_1}-\eqref{eq:N_2} can be directly computed from the definition above,
\al{\label{eq:N_C_coherent}
\braket{N_k^\text{in}} & = \braket{z|a_k^\dagger a_k|z}=|z(\bm k)|^2\\
\braket{C_k^\text{in}} & = \braket{z|a_ka_{-k}|z}=z(\bm k)^2.
}
Therefore, once we divide by the volume $V$ and take the limit $V\to \infty$, we can obtain $\braket{n_k^\text{in}},~\braket{c_k^\text{in}}$ in eqs.\,\eqref{eq:nk_1}-\eqref{eq:nk_2}. However, from the definition in eq.\,\eqref{eq:coherent} we can write the coherent state explicitly as (see eq.\,\eqref{eq:psi_state})
\be
\ket{z}= \sum_n \int\frac{\dd^3k_1}{(2\pi)^3}\cdots\frac{\dd^3k_n}{(2\pi)^3}z_n(\bm k_1,\cdots,\bm k_n)\ket{\bm k_1,\cdots, \bm k_n},
\ee
where each $z_n(\bm k_1,\cdots,\bm k_n)$ is given by
\be
z_n(\bm k_1,\cdots,\bm k_n) = z_0\frac{1}{\sqrt{n!}}\prod_{i=1}^nz(\bm k_i).
\ee
The requirement of normalization $\braket{z|z}=1$ implies
\be
z_0 = \exp\left(-\int\frac{\dd^3k}{(2\pi)^3}|z(\bm k)|^2\right),
\ee
which needs to be finite in order for the state to be well-defined, implying in $|z(\bm k)|^2$ being integrable. As a consequence, $|z(\bm k)|^2$ and $z(\bm k)^2$ cannot possibly scale with the volume $V$ and thus will not contribute to $\braket{n_k^\text{in}},~\braket{c_k^\text{in}}$ in eq.\,\eqref{eq:N_C_coherent}. From these arguments, we conclude that initial states given by coherent states predict the same abundance as in the Bunch--Davies case.

The same argument does not hold for the thermal and squeezed states considered in sec.\,\ref{sec:results}, since in both cases the number of particles produced is extensive, i.e. it scales proportionally to the volume. This is evident from eq.\,\eqref{eq:thermal_distribution}, which describes the number density of particles per comoving momentum $k$ in thermal equilibrium, and from eq.\,\eqref{eq:N_2}, where the volume scaling of the $|\beta_k|^2$ is explicit.

\section{Spectrum for the two-stage inflationary scenario for spin-1 DM}\label{app:spectrum_two_stage_inflation}
We present here the power spectrum $\mathcal{P}_k$ after the field has become non-relativistic. We consider directly the field $u_k(\eta)$, the redefinition of the longitudinal vector component with canonical kinetic term. As discussed in sec.\,\ref{sec:power_spectrum}, at late times this is related to the particle number spectrum by $(k^3/2\pi^2) n_k \simeq \omega_k \mathcal{P}_u$. The computation can be done using the transfer function of eq.\,\eqref{eq:transfer_function_vector}. As in the main text, we have three cases:
\begin{itemize}
\item $k_\star < k_\text{dR}$ (small masses): we obtain
\be
\mathcal{P}_u = \left\{
\begin{aligned}
    & \frac{H_I^2\, a_\star\,k^2}{4\pi^2 m^2\,a} & ~~~ & k < k_\star \\
    & \frac{(a_e \sqrt{H_I\,H_e})^4}{4\pi^2 m\,a\,k} & ~~~ & k_\star < k < k_\text{dR} \\
    & \frac{(a_{\dd\Lambda}\,H_I)^4\,(a_e\,H_e)^4}{4\pi^2 m\,a\,k^5} & ~~~ & k_\text{dR} < k < k_{\dd\Lambda} \\
    & \frac{(a_e\,H_e)^4}{4\pi^2 m\,a\,k} & ~~~ & k_{\dd\Lambda}  < k < k_e
\end{aligned}
\right.
\ee
As we see, we expect a unique peak in the spectrum at $k_\star$;
\item $k_\text{dR} < k_\star < k_{\dd\Lambda}$ (intermediate masses): we have
\be
\mathcal{P}_u = \left\{
\begin{aligned}
    & \frac{H_I^2\, a_\star\,k^2}{4\pi^2 m^2\,a} & ~~~ & k < k_\text{dR} \\
    & \frac{(a_{\dd\Lambda} H_I)^4\, H_e^2\, a_\star}{4\pi^2\,m^2\,a\,k^2} & ~~~ & k_\text{dR} < k< k_\star  \\
    & \frac{(a_{\dd\Lambda}H_I)^4\,(a_e H_e)^4}{4\pi^2 m\,a\,k^5} & ~~~ & k_\star < k < k_{\dd\Lambda} \\
    & \frac{(a_eH_e)^4}{4\pi^2 m\,a\,k} & ~~~ & k_{\dd\Lambda}  < k < k_e
\end{aligned}
\right.
\ee
In this case, there is again a unique peak, but at $k_\text{dR}$;
\item $k_{\dd\Lambda} < k_\star$ (large masses): we obtain
\be
\mathcal{P}_u = \left\{
\begin{aligned}
    & \frac{H_I^2\, a_\star\,k^2}{4\pi^2 m^2\,a} & ~~~ & k < k_\text{dR} \\
    & \frac{(a_{\dd\Lambda} H_I)^4\, H_e^2\, a_\star}{4\pi^2\,m^2\,a\,k^2} & ~~~ & k_\text{dR} < k< k_{\dd\Lambda}  \\
    & \frac{a_\star\,H_e^2\,k^2}{4\pi^2\,m^2\,a} & ~~~ &  k_{\dd\Lambda}< k < k_\star \\
    & \frac{(a_e H_e)^4}{4\pi^2 m\,a\,k} & ~~~ & k_{\dd\Lambda}  < k < k_e
\end{aligned}
\right.
\ee
Here we instead have two peaks, one at $k_\text{dR}$ and another one at $k_\star$.
\end{itemize}
Let us now compute explicitly one of the results to illustrate the method. Take, for instance, the spectrum for small masses ($k_\star < k_\text{dR}$) for modes $k < k_\star$. The mode starts well inside the horizon with Bunch-Davies initial conditions (i.e. power spectrum $k^2/4\pi^2$), which stay constant until horizon exit. The power spectrum then evolves as $a^2$ while outside the horizon. When it finally crosses the comoving Compton wavelength, the power spectrum remains constant until we reach $a_\star$, at which point it start decreasing as $a^{-1}$. We can then write
\be
\frac{k^2}{4\pi^2} \left(\frac{a_\text{Compton}}{a_\text{exit}^{(1)}} \right)^2  \frac{a_\star}{a} = \frac{H_I^2\, a_\star\,k^2}{4\pi^2 m^2\,a},
\ee
where $a_\text{exit}^{(1)} = k/H_I$ is the moment of horizon exit and $a_\text{Compton}  = k/m$ is the moment of Compton crossing. A similar reasoning can be used to compute all other contributions.

\section{Other relevant bounds}\label{app:bounds}

This appendix is dedicated to providing estimates for the isocurvature bounds coming from CMB measurements and from Lyman-$\alpha$ forest data.

\subsection{Isocurvature constraints}\label{app:isocurvature}

In a simplified way, isocurvature perturbations are those that are not spatially aligned with the inhomogeneities of the plasma. These can become sizable for relics originated from non-thermic mechanisms such as the GPP. See for instance ref.\,\cite{Kolb:2023ydq}, and references therein, for a review on the topic. 

Working in comoving gauge for the plasma perturbations, the isocurvature power spectrum produced from a species $X$ coincides with its density contrast power spectrum $\Delta_{\delta_X}^2$, where $\delta_X = \delta\rho_X/\rho_X$ is the density contrast and $\delta\rho_X$ the fluctuations of the energy density $\rho_X$. At late times, after GPP was completed, the two-point correlation between density fluctuations in position space is given by\,\cite{Redi:2022zkt,Garcia:2023qab}
\be
\Delta_{\delta_X}^2=\langle \delta_X(\vec x)\delta_X (\vec x+\vec r)\rangle_{\vec x} = 2\frac{\langle X(\vec x) X(\vec x+\vec r)\rangle_{\vec x}^2}{\langle X^2\rangle^2},
\ee
where the symbol $\langle\cdots\rangle_{\vec x}$ indicates that the average is obtained performing the integration over $\vec x$, and $\langle X^2\rangle \equiv \langle X(\vec x)X(\vec x)\rangle_{\vec x}$ is the the zero-distance correlator that serves as normalization for the power spectrum. The contrast power spectrum is obtained after Fourier-transforming the quantity above:
\be\label{eq:Delta_delta}
\Delta_{\delta_X}^2(k) = \frac{k^3}{\langle X^2\rangle^2}\int_{-1}^1\text{d}\cos\theta\int_0^{\infty}\text{d} p\ \frac{\mathcal{P}_X(p)\mathcal{P}_X(|\vec p - \vec k|)}{p|\vec p - \vec k|^3},
\ee
where we make use of the definition of the power spectrum $\mathcal{P}_X(k)$ of eq.\,\eqref{eq:power_spectrum}.

In all cases studied in this paper, the power spectrum is a broken power-law, peaked at some scale $k_\text{peak}$. Using this property, one can show that
\be\label{eq:Delta_delta_scaling}
\Delta_{\delta_X}^2(k\ll k_\text{peak}) \simeq 2\left(\frac{k}{k_\text{peak}}\right)^3,
\ee
independently of the shape of the power spectrum for low $k$.

The \textit{Planck} collaboration has constrained the isocurvature power spectrum to be\,\cite{Planck:2018jri}
\be\label{eq:isocurvature}
\Delta_{\delta_X}^2(k_\text{CMB}) < \Delta_\text{iso}^2 \simeq 8\times 10^{-11},
\ee
where $k_\text{CMB}\simeq 0.05\,a_0~\text{Mpc}^{-1} \simeq (3.2\times 10^{-40}\,\text{GeV})\, a_0$ is the CMB pivot scale, with $a_0$ the scale factor today. In order to obtain the upper limit above, it is considered that curvature and isocurvature fluctuations are uncorrelated (``axion I" according to ref.\,\cite{Planck:2018jri}), which we assume to hold in our case (see however refs.\,\cite{Chung:2011xd,Chung:2013sla}). Combining eqs.\,\eqref{eq:Delta_delta_scaling} and\,\eqref{eq:isocurvature}, and using entropy density conservation to write $a_0^3 = a_e^3 90 H_e m_\text{PL}^2/(g_{\star s}(T_0) \pi^2 T_\text{RH} T_0^3)$ to write $a_0$ in terms of early universe parameters (with $m_\text{PL}$ the reduced Planck mass, $T_0$ the photon temperature today and $g_{\star s}$ the number of entropy degrees of freedom), we obtain the following lower bound on the peak momentum:
\be\label{eq:kpeak_lowerbound_isocurvature}
\frac{k_\text{peak}}{\text{GeV}} \gtrsim 1.11\times 10^{-6}\, a_e \left(\frac{H_e}{10^{12}\text{GeV}}\right)^{2/3} \left(\frac{10^9\text{GeV}}{T_\text{RH}}\right)^{1/3}.
\ee
In the cases discussed in the text for the vector DM particle, we either have a peak at $k_\star$, at $k_\text{dR}$ or at both momenta, so that we must take $k_\text{peak} = \text{min}(k_\star, k_\text{dR})$. 

When $k_\text{peak} = k_\star$ (usual GPP, thermal initial conditions, two-stage inflation with $m < m_\text{dR}$), we can use eq.\,\eqref{eq:kstar} to write
\be\label{eq:kstar_estimate}
\frac{k_\star}{a_eH_e} \simeq 14 \sqrt{\frac{m}{H_e}}\left(\frac{H_e}{10^{13}~\text{GeV}}\right)^{1/6}\left(\frac{10^{12}~\text{GeV}}{T_\text{RH}}\right)^{1/3}.
\ee
Inserting this value in eq.\,\eqref{eq:kpeak_lowerbound_isocurvature} we find a lower bound on the vector mass, which must be combined with the upper bound $m < m_\text{dR}$. For instance, $H_e=10^{13}$ GeV and $T_\text{RH}=10^{12}$ GeV, masses below $10^{-30}$ GeV are inconsistent with isocurvature constraints.  

In the opposite case, when $k_\text{peak} = k_\text{dR}$ (which applies to the two-stage inflationary case with $m > m_\text{dR}$), we can use
\be
k_\text{dR} = a_\text{dR} H_e = (10^{12}\,\text{GeV}) \,a_e \left(\frac{a_\text{dR}}{a_e}\right) \left(\frac{H_e}{10^{12}\,\text{GeV}}\right).
\ee
Inserting this expression in eq.\,\eqref{eq:kpeak_lowerbound_isocurvature} we obtain the upper bound
\be
\log\left(\frac{a_e}{a_\text{dR}}\right) \lesssim 40 + \frac{1}{3} \log\left(\frac{H_e}{10^{12}\,\text{GeV}} \,\frac{T_\text{RH}}{10^{9}\,\text{GeV}}\right)  .
\ee
In the numerical analysis presented in sec.\,\ref{sec:pre_inflation} this condition is always satisfied, so that no isocurvature limits appear in figs.\,\ref{fig:Spectrum_preinflationary}-\ref{fig:Oh2_preinflationary}.

\subsection{Lyman-$\alpha$ constraints}

Large scale structures in our universe are compatible with DM being cold, i.e. non-relativistic at the onset of structure formation. One way to probe this is through the Lyman-$\alpha$ forest, which consist of spectral lines of quasars and other objects; because the light emitted by them interact with medium at different redshifts, one observes several spectral lines corresponding to the same atomic transition. The Lyman-$\alpha$ forest is thus a probe of how structures are spread in the universe and, indirectly, of the DM velocity. In particular, a too large DM velocity is excluded by Lyman-$\alpha$ limits.

\begin{figure}[t]
\centering
\includegraphics[width=0.48\textwidth]{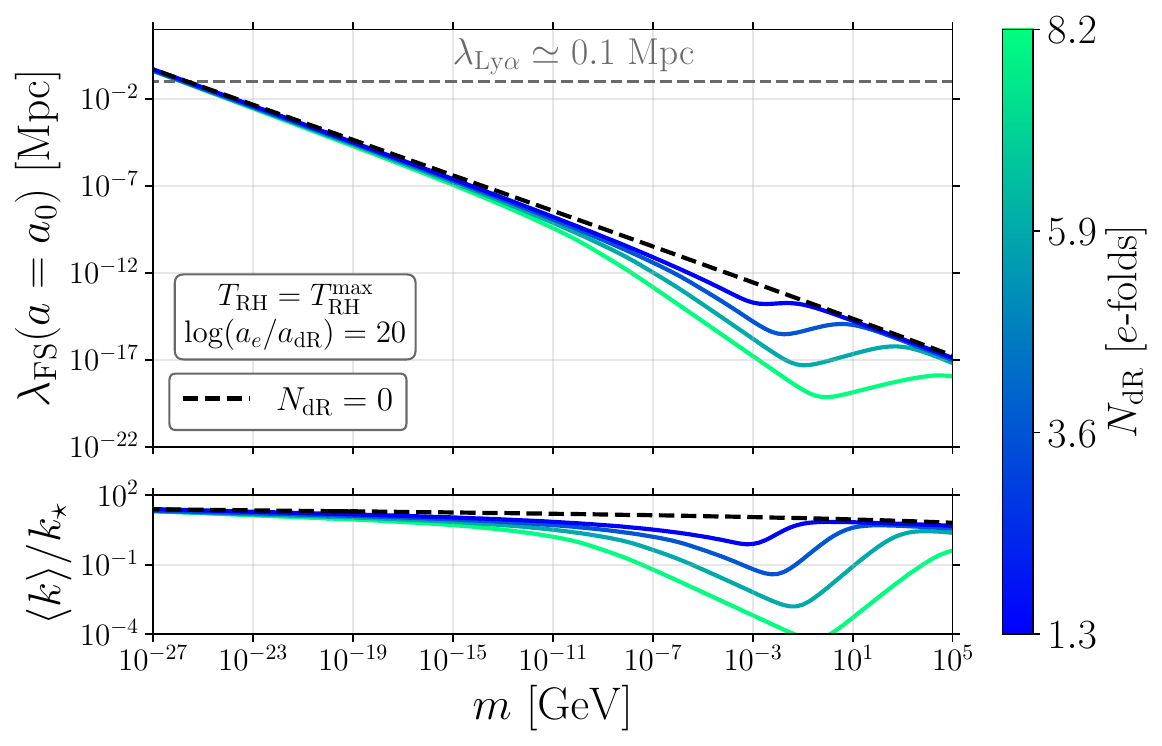}
\caption{Lyman-$\alpha$ constraints for the pre-inflationary scenario, with $T_\text{RH}=T_\text{RH}^\text{max}$ and $\log(a_e/a_\text{dR})=20$. The upper panel shows the mean free-streaming length of eq.\,\eqref{eq:lambda_FS} evaluated today, as a function of the mass of the vector. The lower panel shows the ratio between the average comoving momentum $\braket{k}$, computed according to eq.\,\eqref{eq:k_average}, and $k_\star$, that represents the peak of the spectrum in the usual GPP. Dashed black curves assume inflation with $N_\text{dR}=0$, whereas colors assume a certain amount of $e$-folds of radiation domination $N_\text{dR}$ in the two-stage inflationary scenario, according to the color code.}\label{fig:lyman_alpha}
\end{figure}

We can check if the particles produced via GPP respect this constraint. Consider the average comoving momentum $\braket{k}$ defined by
\be\label{eq:k_average}
\braket{k} \equiv \frac{\int \frac{\dd k}{k}~ k \frac{k^3n_k}{2\pi^2}}{\int \frac{\dd k}{k} ~\frac{k^3n_k}{2\pi^2}},
\ee
with the spectrum defined in eqs.\,\eqref{eq:nk_1}, \eqref{eq:nk_2}.
We can then compute the average velocity of the relic at a certain scale factor $a$ (for $a>a_\star$, with $H(a_\star)=m$) as
\be\label{eq:average_v}
\braket{v(a)} \equiv \frac{\braket{k}/a}{m}.
\ee
If $\braket{v(a)}$ is sizable, it can disrupt the formation of present structures. One way to measure this effect is through the physical free-streaming length at a given time $\lambda_\text{FS}(a)$ (see for instance refs.\,\cite{Boyarsky:2008xj,Kolb:2023ydq,Viel:2013fqw}):
\be\label{eq:lambda_FS}
\lambda_\text{FS}(a)\equiv a\int_{a_\star}^{a}\dd a' \frac{\braket{v(a')}}{{a'}^2H(a')},
\ee
where the integration starts at $a_\star$ because only after that point the relic abundance has formed. Note that this is nothing but the integral of $\braket{v}$ over the conformal time $\dd \eta = \dd t/a = \dd a/a^2H$, meaning that this free-streaming length takes into account the propagation of DM in the expanding universe. In other words, it measures the maximum physical distance the relic can travel at a given scale factor $a$. Therefore, the relics will be subject to the gravitational attraction only for scales larger than $\lambda_\text{FS}(a)$, whereas smaller structures will be inhibited.

The Lyman-$\alpha$ forest data probe structures down to the Mpc scale. Below that, it is still difficult to probe the matter power spectrum, thus allowing for DM to behave as warm for scales $<\mathcal{O}(0.1~\text{Mpc})$\,\cite{Viel:2013fqw}. This means that the free-streaming length evaluated today must be constrained to obey
\be\label{eq:Lyman_alpha_bound}
\lambda_\text{FS}(a=a_0) < \lambda_{\text{Ly}\alpha} \simeq 0.1~\text{Mpc},
\ee
allowing us to put a bound on the relics produced from GPP.

In fig.\,\ref{fig:lyman_alpha} we show our results for the computation of the free-streaming length of eq.\,\eqref{eq:lambda_FS} in the two-stage inflationary case of sec.\,\ref{sec:pre_inflation}. In the standard case with $N_\text{dR}=0$ (dashed black), we see that only for masses $m\lesssim 10^{-26}$ GeV we encounter problems with Lyman-$\alpha$ bounds\,\eqref{eq:Lyman_alpha_bound}, which is also rather insensitive to the presence of a second phase (colored solid lines). Instead, the extra phase of radiation domination prior to standard inflation actually decreases the mean velocity of the DM, as it can be seen from the lower panel. 
In fig.\,\ref{fig:lyman_alpha} we have chosen $T_\text{RH}=T_\text{RH}^\text{max}$ for concreteness, but lower reheating temperatures will not qualitatively change the results just discussed; the smaller $T_\text{RH}$ the smaller $m_\text{dR}$ is, hence the diminish of the free-streaming scale will be shifted to lower masses. We therefore conclude that Lyman-$\alpha$ forest constraints should not be relevant to the parameter space shown in fig.\,\ref{fig:Oh2_preinflationary}.

Similarly, we argue that Lyman-$\alpha$ constraints will not be relevant also for the results presented in sec.\,\ref{sec:thermal}, when considering an initial thermal distribution. As we see from the spectrum in eq.\,\eqref{fig:Spectrum_thermal}, the peak of the distribution is unchanged and therefore the average velocity should be approximately equal to that of the standard GPP case. As a consequence, the same limit $m\lesssim 10^{-26}$ GeV should follow from fig.\,\ref{fig:lyman_alpha} in the thermal case.

\subsection{White-noise}

We highlight that  for particles produced via GPP, another relevant lower bound on their mass $m\lesssim 10^{-28}$ GeV comes from ref.\,\cite{Amin:2022nlh}, which exploits the characteristic white-noise of the density perturbations in order to constraint the free-streaming of the DM. In the cases considered here, this bound is either weaker than other constraints or not relevant in our parameter space.

\pagestyle{plain}
\bibliographystyle{apsrev}
\small
\bibliography{biblio}
\end{document}